\documentstyle[epsfig]{mn}
\newif\ifAMStwofonts

\ifoldfss
  \ifCUPmtlplainloaded \else
    \NewTextAlphabet{textbfit} {cmbxti10} {}
    \NewTextAlphabet{textbfss} {cmssbx10} {}
    \NewMathAlphabet{mathbfit} {cmbxti10} {} 
    \NewMathAlphabet{mathbfss} {cmssbx10} {} 
  \fi
  \ifAMStwofonts
    \ifCUPmtlplainloaded \else
      \NewSymbolFont{upmath} {eurm10}
      \NewSymbolFont{AMSa} {msam10}
      \NewMathSymbol{\upi}     {0}{upmath}{19}
      \NewMathSymbol{\umu}     {0}{upmath}{16}
      \NewMathSymbol{\upartial}{0}{upmath}{40}
      \NewMathSymbol{\leqslant}{3}{AMSa}{36}
      \NewMathSymbol{\geqslant}{3}{AMSa}{3E}

       \let\le=\leqslant
       \let\ge=\geqslant
    \fi
  \fi
\fi 

\ifnfssone
  \newmathalphabet{\mathit}
  \addtoversion{normal}{\mathit}{cmr}{m}{it}
  \addtoversion{bold}{\mathit}{cmr}{bx}{it}
  \newmathalphabet{\mathbfit} 
  \addtoversion{normal}{\mathbfit}{cmr}{bx}{it}
  \addtoversion{bold}{\mathbfit}{cmr}{bx}{it}
  \newmathalphabet{\mathbfss} 
  \addtoversion{normal}{\mathbfss}{cmss}{bx}{n}
  \addtoversion{bold}{\mathbfss}{cmss}{bx}{n}
  \ifAMStwofonts
    \ifCUPmtlplainloaded \else
      \UseAMStwoboldmath
      \makeatletter
      \new@mathgroup\upmath@group
      \define@mathgroup\mv@normal\upmath@group{eur}{m}{n}
      \define@mathgroup\mv@bold\upmath@group{eur}{b}{n}
      \edef\UPM{\hexnumber\upmath@group}
      \new@mathgroup\amsa@group
      \define@mathgroup\mv@normal\amsa@group{msa}{m}{n}
      \define@mathgroup\mv@bold\amsa@group{msa}{m}{n}
      \edef\AMSa{\hexnumber\amsa@group}
      \makeatother
      \mathchardef\upi="0\UPM19
      \mathchardef\umu="0\UPM16
      \mathchardef\upartial="0\UPM40
      \mathchardef\leqslant="3\AMSa36
      \mathchardef\geqslant="3\AMSa3E

       \let\le=\leqslant
       \let\ge=\geqslant
    \fi
  \fi
\fi 

\ifnfsstwo
  \DeclareMathAlphabet{\mathbfit}{OT1}{cmr}{bx}{it}
  \SetMathAlphabet\mathbfit{bold}{OT1}{cmr}{bx}{it}
  \DeclareMathAlphabet{\mathbfss}{OT1}{cmss}{bx}{n}
  \SetMathAlphabet\mathbfss{bold}{OT1}{cmss}{bx}{n}
  \ifAMStwofonts
    \ifCUPmtlplainloaded \else
      \DeclareSymbolFont{UPM}{U}{eur}{m}{n}
      \SetSymbolFont{UPM}{bold}{U}{eur}{b}{n}
      \DeclareSymbolFont{AMSa}{U}{msa}{m}{n}
      \DeclareMathSymbol{\upi}{0}{UPM}{"19}
      \DeclareMathSymbol{\umu}{0}{UPM}{"16}
      \DeclareMathSymbol{\upartial}{0}{UPM}{"40}
      \DeclareMathSymbol{\leqslant}{3}{AMSa}{"36}
      \DeclareMathSymbol{\geqslant}{3}{AMSa}{"3E}

       \let\le=\leqslant
       \let\ge=\geqslant
    \fi
  \fi
\fi 

\ifCUPmtlplainloaded \else
  \ifAMStwofonts \else 
    \def\upi{\pi}
    \def\umu{\mu}
    \def\upartial{\partial}
  \fi
\fi

\title{Chemical evolution of local galaxies in a hierarchical model}
\author[F. Calura, N. Menci]
       {F. Calura$^{1,2}$\thanks{E-mail: fcalura@oats.inaf.it}, N. Menci$^{3}$\\
        (1) Dipartimento di Astronomia-Universit\'a di Trieste, Via G.B. Tiepolo
	11, 34131 Trieste, Italy\\
        (2) INAF, Osservatorio Astronomico di Trieste, Via G.B. Tiepolo
	11, 34131 Trieste, Italy\\
	(3) INAF, Osservatorio Astronomico di Roma, via Frascati 33, I-00040 Monteporzio, Italy\\
	 }
	
\date{Accepted ---- .
      Received ---- ;
      in original form ----}

\pagerange{\pageref{firstpage}--\pageref{lastpage}}
\pubyear{----}

\begin{document}

\maketitle

\label{firstpage}
\begin{abstract}
We investigate the chemical properties of local galaxies within a cosmological framework 
in the hierarchical picture of galaxy formation. 
To this aim, we use a hierarchical semi-analytic model which includes 
the contribution from (i) low and intermediate mass stars, 
relevant producers of some important elements, such as C and N; 
(ii) type Ia Supernovae (SNe), for which a continuous delay-time distribution is assumed and which are important 
for the production of Fe and (iii) massive stars, dying as core-collapse type II SNe and which produce 
the $\alpha-$elements. In this way, we can study abundances  
for a large set of chemical elements and in various galactic types, 
comparing our predictions with available 
observations in the Milky Way (MW), in local dwarf galaxies and in local ellipticals.\\
For Milky-Way-like galaxies, we can successfully reproduce the major observational constraints, i.e. 
the [O-Fe] vs [Fe/H] relation observed in disc stars and the stellar metallicity distribution (SMD). 
For dwarf galaxies, the stellar metallicity vs mass relation is 
reproduced by assuming that a substantial fraction of the heavy elements is lost through 
metal-enhanced outflows and a type Ia SN realization probability lower than the one of MW-like galaxies.
The predicted abundance ratios for dwarf galaxies are comparable to observations derived locally 
for dwarf spheroidals. The stellar metallicity distributions predicted for dwarf galaxies 
are in agreement with the local observations. 
We predict a substantial presence of extremely 
low metallicity stars [Fe/H]$<$-2.5, which have been recently observed in ultra-faint dwarf galaxies.  
In ellipticals, 
the observations indicate 
higher [$\alpha$/Fe] values in larger galaxies. 
Several previous attempts to model the [$\alpha$/Fe] vs $\sigma$ in ellipticals based 
on $\Lambda-$cold dark matter galaxy formation models predicted an anti-correlation between 
[$\alpha$/Fe] vs $\sigma$, indicating too much extended star formation histories in high-mass galaxies. 
Our results computed with a standard Salpeter initial mass function (IMF) indicate a flat [$\alpha$/Fe] vs relation. 
However, we suggest a possible solution to this problem and show how, by assuming a star formation-dependent IMF 
with a slope $x=1.35$ in systems with star formation rates 
$< 100 M_{\odot}/yr$ and slightly flatter (i.e. with $x=1$) in object with stronger star formation, 
the observed correlation between [$\alpha$/Fe] and  $\sigma$
can be accounted for on a large velocity dispersion range. 
Fundamental roles are played also by interaction-triggered starbursts and AGN feedback. 
Finally, a SF-dependent IMF seems necessary  also to reproduce the 
stellar metallicity-$\sigma$ relation observed in local early-type galaxies. 
\end{abstract} 


\section{Introduction} 
In the last few years, the power of high-resolution 
spectrographs has made available a 
large amount of stellar and interstellar abundances for 
various chemical species. 
Nowadays, large catalogues of 
stellar abundances are accessible for the Milky Way galaxy 
and for the dwarf galaxies of the Local Group. Gas-phase 
abundances have been derived up to high-redshift ($z>3$), both 
by analysing emission lines and absorption lines, present in the spectra of distant 
Quasars. This large set of data provides us with valuable constraints for galactic chemical 
evolution studies. Of particular interest is the study of chemical evolution 
by means of ``ab-initio'' galaxy formation models. \\
The study of galactic chemical evolution within a cosmological framework 
is of particular importance, since it provides us with crucial information on 
the star formation history (SFH) of galaxies, 
on the ages of the stellar populations and on the gas accretion and outflow histories. 
Chemical evolution offers a way to constrain the main parameters driving  
all of these processes, which are of primary 
interest for galaxy formation theories.  \\
So far, a few theoretical studies of hierarchical galaxy formation 
have payed attention to the problem of galactic 
chemical evolution.  
In a pioneeristic paper, Thomas (1999) considered 
star formation histories from a hierarchical 
semi-analytic galaxy formation model, taking into account for   
the first time the rates of type Ia SNe. This attempt allowed him to compute 
the evolution of the abundance ratio between elements formed on different timescales, 
i.e. O, synthesised by type II Supernovae (SNe)  on timescales less than 0.03 Gyr, and 
Fe, produced mainly by type Ia SNe on timescales ranging from 0.03 Gyr up to one Hubble time. 
Other important works on chemical evolution in the hierarchical framework are the ones 
by Nagashima et al. (2005), and Nagashima \& Okamoto (2006), 
where the chemical evolution of various elements produced by type Ia and type II SNe  
is computed. 
However, none of the works quoted above takes into account 
the role of low and intermediate mass stars (LIMS hereinafter), i.e. all the stars with masses 
$0.8 \le m/M_{\odot} \le 8$. These stars are of fundamental importance in chemical evolution 
studies since they incorporate an important fraction of the metals 
and during the last stages of their evolution, i.e. during the planetary nebula phase, 
they may restore variable 
amounts of non-processed heavy elements incorporated at their birth and 
produced by stars of previous generations, giving a significant contribution 
to the heavy element pollution of the interstellar medium (ISM). 
Furthermore, low and intermediate mass stars are relevant producers of 
C and N (Van den Hoeck and Groenewegen 1997), whose abundances can not be 
properly assessed by any chemical evolution model which does not take into account 
in detail the evolution of LIMS.  
In this paper, with a hierarchical semi-analytical model (SAM)
we study the chemical evolution of these  heavy elements, taking into account also the 
contribution from LIMS. 
Our main aim is providing chemical evolution predictions computed by means of  
star formation (SF) and gas accretion histories derived from an 
ab-initio galaxy formation model.  
For the first time, by means of a SAM for galaxy formation, 
we study the evolution of the abundances for a large set of chemical elements 
produced by stars of various masses, ending their lives on various timescales. 
This paper is mainly focused on the chemical abundances in local galaxies. 
We compare our predictions to abundances observed 
in various galactic environments, such as in 
the Milky-Way disc, local dwarf galaxies and local early-type galaxies. 
Future papers will be devoted to the study of chemical abundances at high redshift. 
This paper is organised as follows. 
In Sect. 2, we present our theoretical instruments, i.e. the hierarchical semi-analytical 
galaxy formation model and the main chemical evolution equations. In Sect. 3, 
we present our results. Finally, in Sect. 4 we discuss the main implications of our results and draw some conclusions.

\section{The model }
In this paper, we start from the star formation histories of 
galaxies computed by means of a semi-analytic model of galaxy formation. 
For each star formation history, 
we compute the chemical evolution a posteriori by means 
of detailed chemical evolution equations. In this section, we briefly describe the SAM 
used in this work and the methods used to compute the evolution 
of the chemical abundances. 

\subsection{The galaxy formation model} 
\label{SAM}
We derive the star formation histories of galaxies and of their progenitors from the semi-analytic model by Menci et al. 
(2006, 2008); we recall here its key features. Galaxy formation and evolution is driven by the collapse and growth of dark
matter (DM) halos, which originate by gravitational instability of  overdense
regions in the primordial DM density field. This is taken to be a random,
Gaussian  density field with Cold Dark Matter (CDM) power spectrum within the
''concordance cosmology" (Spergel et al. 2006) for which we adopt round
parameters  $\Omega_{\Lambda}=0.7$, $\Omega_{0}=0.3$, baryonic density
$\Omega_b=0.04$ and Hubble constant (in units of 100 km/s/Mpc) $h=0.7$. The
normalisation of the spectrum is taken to be $\sigma_8=0.9$ in terms of the variance
of the field smoothed over regions of 8 $h^{-1}$ Mpc.

As  cosmic time increases, larger and larger regions of the density field
collapse, and  eventually lead to the formation of groups and clusters of
galaxies, which grow by merging with mass and redshift dependent rates
provided by the Extended Press \& Schechter formalism (see
Bond et al. 1991; Lacey \& Cole 1993).  The clumps included into larger DM halos
may survive as satellites, or merge to form larger galaxies due to binary
aggregations,  or coalesce into the central dominant galaxy due to dynamical
friction; these processes take place on timescales that grow longer over
cosmic time, so the number of satellite galaxies increases as the DM host halos
grow from groups to clusters. All the above processes are implemented in our
model  as described in detail in Menci et al. (2005, 2006), based on canonical
prescriptions of semi-analytic modeling.

\subsubsection{Star Formation and SN feedback}
\label{SN}
The radiative gas cooling, the ensuing star formation and the
Supernova events with the associated feedback occurring  in the growing
DM halos (with mass $M$ and circular velocity $v$) are described in
Menci et al. (2005). The cooled gas with mass
$M_g$ settles into a rotationally supported disk with radius $r_d$
(typically ranging from $1$ to $5 $ kpc), rotation velocity $v_d$
and dynamical time $t_d=r_d/v_d$. The gas gradually condenses  into
stars and the stellar  feedback
returns part of the cooled gas to the hot gas phase
with mass $M_h$ at the virial temperature of the halo. 
 As for the star formation, we assume the canonical Schmidt form $
\psi = M_{g}/(q\,\tau_d)$, where $\tau_d\equiv r_d/v_d$ and $q$ is fixed by the Kennicutt (1998) law. 
At each time step, the mass $\Delta m_h$ returned from the cold
gas content of the disk to the hot gas phase due to Supernovae (SNe) activity is
estimated from canonical energy balance arguments (Kauffman 1996, Kauffmann \&
Charlot 1998; see also Dekel \& Birnboim 2006) as 
\begin{equation}
\Delta m_h=E_{SN}\,\epsilon_0\,\eta_0\,\Delta m_*/v_c^{2} 
\label{m_h}
\end{equation}
where $\Delta m_*$ is the
mass of stars formed in the timestep, $\eta\approx 3-5\cdot 10^{-3}/M_{\odot}$
is the number of SNe per unit solar mass (depending on the assumed
IMF), $E_{SN}=10^{51}\,{\rm ergs}$ is the energy of ejecta of each SN, and $v_c$
is the circular velocity of the galactic halo; $\epsilon_0=0.01-0.5$ is the
efficiency of the energy transfer to the cold interstellar gas. The
above mass $\Delta m_h$ is made available for cooling at the next timestep. The
model free parameters $q=30$ and $\epsilon_0=0.1$ are chosen as to match
the local B-band luminosity function and the Tully-Fisher relation adopting a Salpeter IMF. \\
Note that our simple modelling for SNae feedback provides a  very good fit to the observed
correlations of outflow velocities with galactic properties (like the circular
velocity or the star formation rate). Although an accurate estimate of the outflow velocity
would require a detailed treatment of the gas kinematics (including also the physics of OB
associations and the dynamics of of superbubbles, see Ferrara et al. 2000; Veilleux et al. 2005)
and is beyond the aim of the present paper, its asymptotic value $V_{ouflow}$ can be estimated from the energetic balance
between the total energy deposition rate by SNe and the rate of kynethic energy loss as (Veilleux et al. 2005)
$ \Delta m_h\,V_{max}^{2} = E_{SN}\,\epsilon_0\,\eta_0\,\psi$.
A comparison with our assumed expression for the expelled gas mass $\Delta m_h$ given above
implies $V_{max}\approx v_c$ (or equivalently a mass loading factor $\dot m_h/\psi\propto v_c^{-2}$), a relation which
is in very good agreement with the relation observed in local starbursts (Martin 2005; Rupke et al. 2005; Veilleux et al. 2005).

\subsubsection{AGN feedback}
\label{AGN}
The model also includes a treatment of the growth of supermassive
black holes at the centre of galaxies by interaction-triggered inflow
of cold gas, following the physical model of Cavaliere \& Vittorini
(2000). 
Our SAM includes a detailed treatment of feedback from Active Galactic Nuclei (AGN),
which acts only during the active AGN phases of each galaxy (hence
for a minor fraction $\sim 10^{-2}$ of the galaxy lifetime).
This is assumed to stem from the fast winds with velocity up to
$10^{-1}c$ observed in the central regions of AGNs
(Weymann 1981; Turnshek et al. 1988;  Risaliti et al. 2005);
these are usually though to originate from the acceleration of disk outflows due to the AGN
radiation field (Proga 2007 and references therein, Begelman 2003).
These supersonic outflows compress the gas into a blast wave terminated by
a leading shock front, which moves outwards with a lower but still
supersonic speed and sweeps out the surrounding medium. Eventually,
this is expelled from the galaxy.
Quantitatively, the energy injected into the galactic gas
in such inner regions is taken to be proportional to the energy radiated by the
AGN, $\Delta E = \epsilon_{AGN}\,\eta\,c^2\,\Delta m_{acc}$.
the value of the energy feedback efficiency for coupling with
the surrounding gas is taken as
$\epsilon_{AGN}=5 \cdot  10^{-2}$, consistent with the values required to
match the X-ray properties of the ICM in clusters of galaxies (see
Cavaliere, Lapi \& Menci 2002). This is also consistent with the
observations of wind speeds up to $v_w\approx 0.1\,c$ in the central
regions, that yield $\epsilon_{AGN}\approx v_w/2c\approx 0.05$ by
momentum conservation between photons and particles
(see Chartas et al. 2002, Pounds et al. 2003); this value has been also adopted
in a number of simulations (e.g., Di
Matteo, Springel \& Hernquist 2005) and semi-analytic models
of galaxy formation (e.g., Menci et al. 2006).
The transport of the above energy in the galaxy and its effect on the distribution of the
galactic gas is computed in detail by describing the expansion of the blast wave
solving the corresponding hydrodynamical equations; these include the effects not
only of initial density gradient, but also those of upstream
pressure and DM gravity (Lapi, Cavaliere, Menci 2005). The
solutions show in detail how the
perturbed gas is confined to an expanding shell bounded by an outer
shock at the radius $R_s(t)$ which sweeps out the gas surrounding
the AGN. From the resulting shock expansion law $R_s(t)$
the amount of expelled gas can be computed (for different
AGN luminosities and galactic properties), as well as the
final galactic gas distribution, as described in detail in Menci et al. (2008) 
We refer to Menci et al. (2005-2008) for details and for the comparison of the 
model results with observations concerning the cosmological evolution of 
both the galaxy and the AGN population.

\subsubsection{Starbursts triggered by galaxy interactions}
An additional
channel for star formation implemented in the model is provided by
interaction-driven starbursts, triggered not only by merging but
also by fly-by events between galaxies; such a star formation mode
provides an important contribution to the early formation of stars
in massive galaxies, as described in detail in Menci et al. (2004,
2005). 
The galaxy interaction rate is given by $\tau_r^{-1}=n_T\,\Sigma (r_t,v,V_{rel})\,V_{rel}$,
where $n_T$ is the number of galaxies hosted in a given dark matter halo,
$\Sigma$ is the cross section for grazing (i.e. at distances closer than the
galaxy tidal radius) encounters, computed as in Menci et al. (2003)
from the orbital parameters (the impact parameter $b$, the radius of the host dark matter halo)
for each galaxy in our Monte Carlo simulations; the relative velocity $V_{rel}$ is computed
from the velocity dispersion of the dark matter halo hosting the interacting galaxies.
With a probability given by the above equation in each time step, a galaxy is considered to be in a starburst phase.
In such a case, the cold gas fraction converted into stars during the burst is
$f_{acc}\approx
\Big|{\Delta j\over j}\Big|=
10^{-1}\Big\langle {m'\over m}\,{r_d\over b}\,{v_d\over V_{rel}}\Big\rangle\, .
$
The fraction of cold gas accreted by the black hole in an interaction event is here computed
in terms the variation $\Delta j$ of the specific angular momentum $j\approx
Gm/v_d$, induced by the grazing encounter. 
Here $m'$ is the mass of the partner galaxy in the
interaction, and the average runs over the probability of finding such a galaxy
in the same halo where the galaxy with mass $m$ is located. The duration of the burst is
given by the interaction timescale $r_d/V_{rel}$. With such a description, a fraction of
gas $\approx 10 \%$ is impulsively converted into stars during minor bursts with $m'/m\approx 0.1$,
while in the rarer major bursts with $m'/m\approx 1$ the fraction of gas converted into stars can approach
$100\%$.

\subsection{Galactic chemical evolution}
\label{chem}
In this paper, we model chemical evolution following the approach of 
Matteucci \& Greggio (1986). 
We follow the time evolution of the abundances for the following set of chemical elements: H, He, C, N, O, Mg, Si, Fe. 
This is one of the major novelties of this work, since so far such a large set of chemical elements has never been 
considered in semi analytic models for galaxy formation.\\
For any chemical element $i$, the variation of its mass 
in the interstellar gas per unit time is:
\begin{eqnarray}
\frac{dm_{i}}{dt} =-X_{i}(t)\psi(t)+ R_{LIMS, i}(t)
 + R_{Ia,i}(t) + \nonumber \\ R_{II,i}(t)+I_i(t)+O_i(t)      
\label{chemeq}
\end{eqnarray}
where $X_{i}$ is the mass fraction in the gas for the element $i$ 
at the time $t$, $\psi(t)$ is the star formation rate (SFR) of the selected galaxy. 
The terms $R_{LIMS, i}(t)$, $R_{ia, i}(t)$ and  $R_{II,i}(t)$ are 
the rates of production of the element $i$ from 
stars of low and intermediate mass, type Ia SNe and type II SNe, respectively.\\ 
Finally, the terms $I_i(t)$ and $O_i(t)$ take into account possible mass increments and outflows of mass in the form of the element $i$, 
respectively. 
The mass increment may occur by means 
of three processes: (i) cooling of hot gas; (ii) by means of 
cool gas accretion or (iii) merging with other galaxies,  
whereas an outflow can occur owing to a 
galactic wind. 
In the remainder of the paper, when we will use the expression ``Infall'', 
we will intend actually the increment of gas, which 
can occur by means of any of the three processes mentioned above. 
The methods used to compute 
the accretion and outflow histories of the galaxies will be described in Sect.~\ref{inf_out}. \\
Stars of low and intermediate mass are those  with initial masses $0.8\le m/M_{\odot} \le 8$. 
Their production rate for the element $i$ is given by: 
\begin{equation}
R_{LIMS, i}(t) = \int_{0.8}^{M(t)} \psi(t-\tau_m) m_{LIMS,i}(m) \phi(m) dm
\end{equation}
where $M(t)$ is the turnoff mass at the time $t$, 
$\tau_{m}$ is the lifetime 
of the star of mass $m$, 
$\phi(m)$ is the initial mass function (IMF) and $m_{LIMS,i}(m)$ is the total mass 
 in the form 
of the the element $i$
that a low or intermediate mass star restores into the ISM. \\
Unless otherwise stated, for the initial mass function, we assume a Salpeter (1995) law, 
given by $\phi(m)=C \cdot m^{-(1+x)}$ with $x=1.35$. The constant $C$ 
is determined by the condition
\begin{equation}
\int_{0.1}^{100 }m \, \phi(m) \, dm =1
\end{equation}
i.e. from the normalisation to 1 of the IMF by mass. 
The quantities $m_{LIMS,i}(m)$ are taken from Van den Hoeck\& Groenewegen (1997), 
who computed theoretical yields for various elements and for stars of low and intermediate mass as a function 
of the initial metallicity $Z$. The yield $p_i(m)$ is defined as the fraction of the initial mass 
that a star transforms into the chemical element $i$ and restores into the ISM. 
The quantity $m_{LIMS,i}(m)$ is computed from the yield according to 
\begin{equation}
m_{LIMS,i}(m)=X_{i}(t-\tau_m) m_{ej}(m) + m \, p_i(m)
\end{equation}
where $X_{i}(t-\tau_m)$ and  $m_{ej}(m)$ are
the original abundance of the element $i$  computed at the time $t-\tau_m$, i.e. when the star of mass $m$ formed, 
and the total ejected mass for a star of initial 
mass $m$, respectively. \\
For the element $i$, the production rate by type Ia SNe 
is  $R_{ia, i}(t) = m_{Ia,i} R_{Ia}(t)$. 
$R_{Ia}(t)$ is the type Ia supernova (SN) rate at the time $t$, computed according to: 
\begin{equation}
R_{Ia}(t)=k_{\alpha} A_{Ia} \int^{min(t, \tau_x)}_{\tau_i}{ \psi(t-\tau) 
DTD(\tau) d \tau}, 
\label{SNRIa}
\end{equation}
(Greggio 2005; Matteucci et al. 2006). 
The quantity $k_{\alpha}$ is the number of stars per unit mass in a stellar 
generation, given by:
\begin{equation}
k_{\alpha}=\int_{m_l}^{M_L}\phi(m)dm
\end{equation}
where $m_l=0.1 M_{\odot}$ and $M_L=100 M_{\odot}$, whereas  
$ A_{Ia}$ is the realization probability for type Ia SNe and represents  the  
fraction of binary systems which may give rise to type Ia SNe, assumed to be 
constant in time. {This quantity is uncertain (Maoz 2008) and is likely to depend on the environment
(Matteucci et al. 2006). 
In chemical evolution studies, $ A_{Ia}$ is  basically treated as a free parameter, 
tuned in order to reproduce 
the type Ia SN rate in local galaxies (see Matteucci et al. 2006; Calura \& Matteucci 2006). We will test the 
impact of this parameter on some of the results described in this paper.\\
The function $DTD(\tau)$ is the 
delay time distribution. We 
assume that Type Ia SNe originate from 
the explosion of a C/O white dwarf (WD) in a close binary system, 
where the companion is either a Red Giant or a Main-Sequence star. 
This is the  single-degenerate (SD) model  of Whelan \& Iben (1973). 
The $DTD(\tau)$ is from Matteucci et al. (2006 see also Matteucci \& Recchi 2001, Valiante et al. 2009).  \\
$m_{Ia,i}$ is the mass that a type Ia SN synthesises and ejects into the ISM in the form of the element $i$. 
These quantities are taken from Iwamoto et al. (1999). \\
We assume that all stars with mass $m\ge 8 M_{\odot}$ explode as type II SNe. The 
rate of production of the element $i$ from type II SNe  is: 
\begin{eqnarray}
R_{II,i}(t)=\int_{8}^{100} \psi(t-\tau_m) m_{II,i} (m) \phi(m) dm \nonumber \\ \simeq \psi(t) \int_{8}^{100} m_{II,i} (m) \phi(m) dm
\end{eqnarray}
the above simplification means to assume the instantaneous recycling approximation 
to treat chemical enrichment from massive stars, i.e. to assume that all stars with masses 
$m\ge 8 M_{\odot}$ have very small lifetimes. 
This assumption is  motivated by the fact that 
the timestep we use to compute chemical evolution is of $\sim 0.015$ Gyr, 
i.e. half the lifetime of a massive star of $8M_{\odot}$. With such a timestep, 
the effects of the lifetimes of single massive stars of various masses can not be appreciated. \\
Chemical enrichment from type II SNe is computed by means of the quantity $m_{II,i}$
\begin{equation}
m_{II,i} (m)= m_{II,new,i}+m_{ej} \cdot X_i (t)
\label{m_II}
\end{equation}
where 
\begin{equation}
m_{II,new,i} = p_i(m) m   
\end{equation}
The quantity $p_i(m)$ is the fraction of the mass $m$ tranformed into  
the element $i$ and is commonly known as the yield.  
The second term on the right side of eq.~\ref{m_II} is   
the mass in the form of the element $i$ already present when the star formed: $m_{ej}$ 
is the total mass ejected by a star of mass $m$ and 
$X_i$ is the interstellar mass fraction of the element $i$ at the time $t$. 
For massive stars, we adopt the metallicity-dependent yields of Woosley \& Weaver (1995). \\

\subsection{Determining the past star formation, accretion and the outflow history}
\label{inf_out}
 For any galaxy 
at redshift $z\sim 0$, it is possible to reconstruct its star formation rate $\psi(t)$ and its accretion 
and outflow histories on the basis of its past merging history. 
We have already described the possible gas accretion processes in Sect.~\ref{chem}. 
The outflows may originate owing to the constribution of SN explosions and  of the AGN
feedback. 
As described in Section ~\ref{SN} and ~\ref{AGN}, 
both mechanisms inject energy into the ISM and cause gas to move from the cold phase to the hot phase. 
The hot gas may eventually be ejected if its thermal energy is larger than 
its binding energy, due to the gravitational well of both the dark matter and the baryons.
For a particular galaxy, its star formation rate at a past time $t$ is 
given by 
\begin{equation}
\psi(t)=\Sigma_j \psi_j(t)
\end{equation}
where  $\psi_j(t)$ is the SFR of the  $j$-th progenitor of the galaxy at the time $t$. 
This formula allows us to take into account the complex merging history of galaxies,  
playing an important role in chemical evolution. 
In a similar way, at any past time $t$, 
for any galaxy we are able to determine the total cold gas  mass $M_g(t)$. 
If $\Delta_{gas} (t) = M_g(t+dt) - M_g(t) - \Sigma_i R_{*,i} \cdot dt +\psi(t) dt$ 
is the difference in the total gas mass at two 
following timesteps minus the total mass restored by the stars in $dt$ plus the mass 
ending in stars,  
 we use this quantity to 
compute the total ``increment'' and outflow rate at any timestep. 
If $\Delta_{gas}(t)/M_{\odot}>0 $ then we assume that at the time $t$
\begin{eqnarray}
I(t)=\Delta_{gas}/dt  \, \, \, \, \, \, \, \, \, ;  \, \, \, \, \, \, \, \, \,  O(t)=0
\end{eqnarray}
otherwise, 
\begin{eqnarray}
I(t)=0  \, \, \, \, \, \, \, \, \, ;  \, \, \, \, \, \, \, \, \,  O(t)=\Delta_{gas}(t)/dt
\end{eqnarray}
By means of these quantities, 
we are able to solve the chemical evolution equations described in Sect.~\ref{chem}. 
For any chemical element $i$, $I_i(t) = X_{i} (t) \, I(t)$ in case of merging or 
$I_i(t) = X_{i,p} (t) \, I(t)$ in case of infall of pristine gas, where $X_{i,p}$ is the primordial mass fraction 
for the element $i$.  

In The SAM, the metal-enriched gas expelled from satellite galaxies residing 
in a common DM halo is diluted by the hot gas of  the halo, whose amount is much larger, generally by a factor 100, 
than the amount of the interstellar cold gas. 
For this reason, the metallicity of the gas 
which may eventually fall back from the hot halo  onto a galaxy, which in general is the central galaxy of the halo,  may be considered negligible. This allows us to  assume that the accreted gas has a primordial composition, i.e. 
$X_{H,p}=0.75 $, $X_{He,p}=0.25 $ and for any other element, we assume  $X_{i,p}=0$.
On the other hand, for any element, unless otherwise stated, we assume $O_i(t) = X_{i} (t) \, O(t)$, i.e. the chemical composition of 
the outflow is the same as the one of the cold ISM.

\section{Results}
In this section, we describe the results 
obtained for various types of galactic systems. 
These results concern mainly local galaxies and are compared to 
observational constraints obtained in the Milky Way, in local dwarf galaxies 
and in local ellipticals. A study of the chemical evolution of distant 
galaxies is the subject for future work. \\
At $z\sim 0$, from the catalogue of galaxies 
obtained by means of the SAM of Menci et al. (2008), 
we select galaxies on the basis of various criteria. 
The selection criteria are similar to those adopted for the observations, and are 
based on the combined use of various available  
quantities, such as the colour, star formation rate or the stellar mass. 
In the following sections, we describe how we select 
galaxies of various types  and our comparison with observational 
data from the literature. 

\begin{figure*}
\centering
\vspace{0.001cm}
\epsfig{file=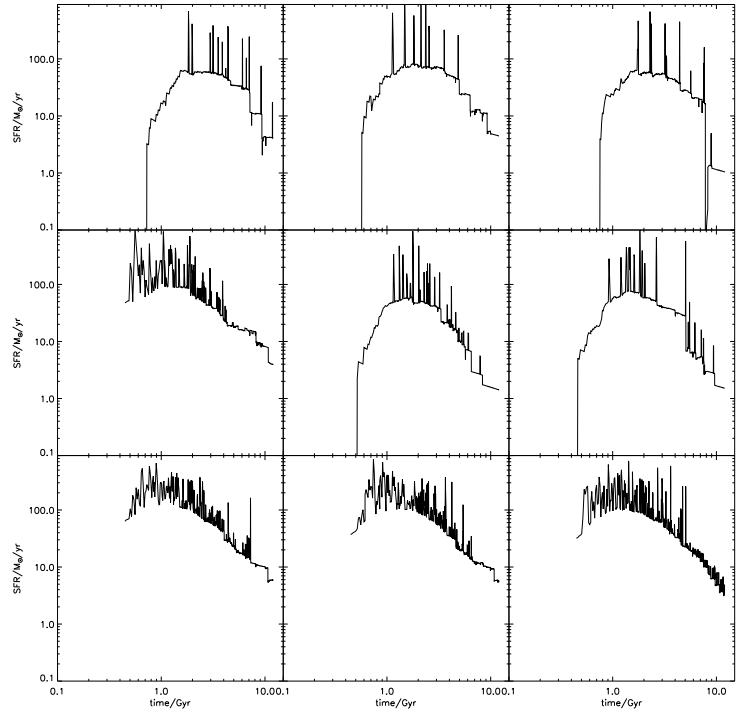,height=10cm,width=10cm}
\caption{ Star Formation histories of a few MW-like galaxies selected according the criteria described in 
Sect.~\ref{MW}.}
\label{sfr}
\end{figure*}
\begin{figure*}
\centering
\vspace{0.001cm}
\epsfig{file=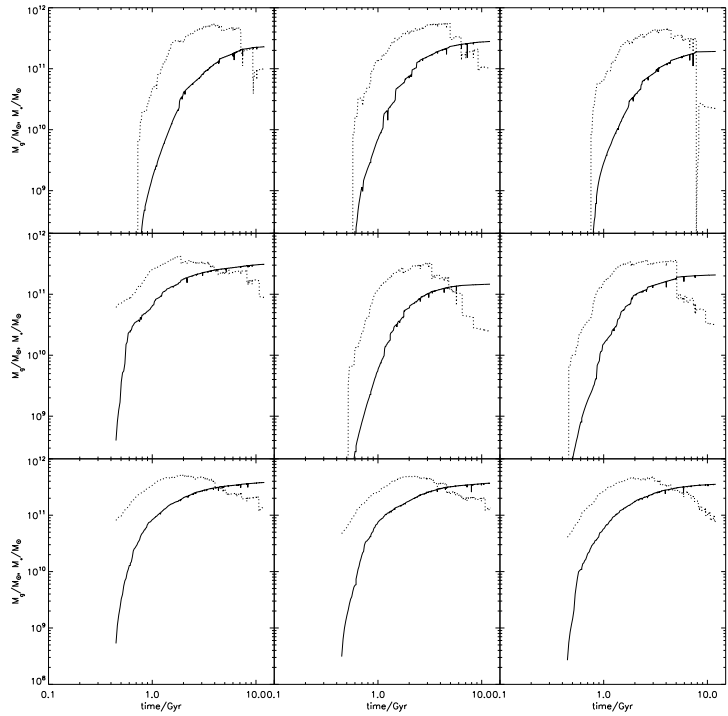,height=10cm,width=10cm}
\caption{Evolution of the cumulative stellar masses (solid lines) and gas masses (dotted lines) for 
a few MW-like galaxies selected according the criteria described in 
Sect.~\ref{MW}.}
\label{gas}
\end{figure*}

\begin{figure*}
\centering
\vspace{0.001cm}
\epsfig{file=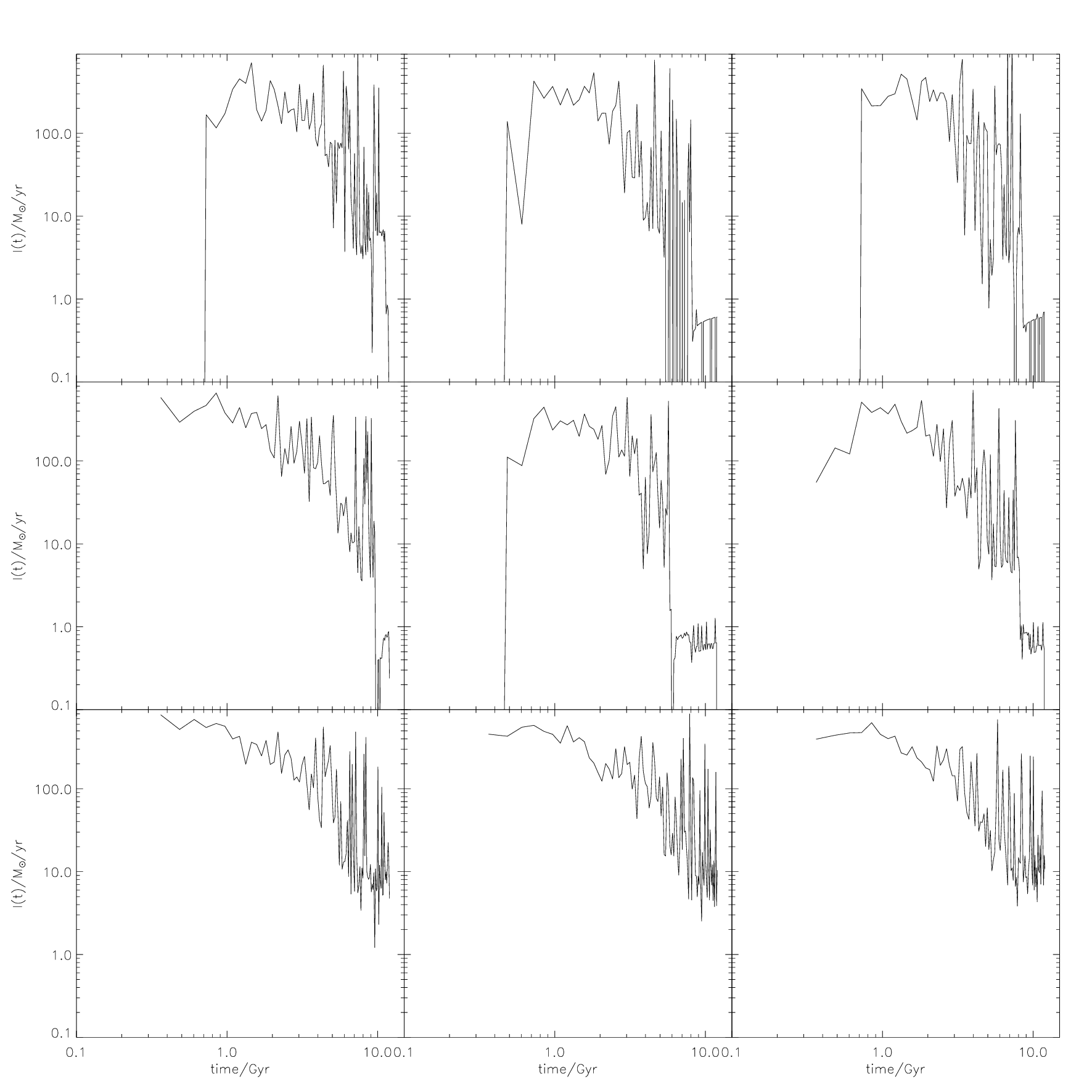,height=10cm,width=10cm}
\caption{Time evolution of the increment rate $I(t)$ for 
a few MW-like galaxies selected according the criteria described in 
Sect.~\ref{MW}.}
\label{infall}
\end{figure*}

\subsection{Chemical abundances in Milky-Way like galaxies}
\label{MW}
The first important 
step of our study is to test our predictions for Milky-Way (hereinafter MW) like 
galaxies. 
Recently, De Rossi et al. (2009) published a detailed study of MW-like galaxies extracted 
from the Millennium Simulation catalogue. 
Following De Rossi et al. (2009), we define MW-like galaxies as those 
with circular velocity $V_c$ in the range 
$200 < V_c/km \, s^{−-1} < 240 $  and cold 
gas fractions $f_c=M_{g}/(M_{g}+M_{*})$ in the range 
$0.1 \le f_c \le 0.3$. \\
By means of these criteria, for the local number density of MW-like galaxies, 
we obtain  $\sim 1 \cdot 10^{-4} Mpc^{-3}$, consistent with 
De Rossi et al. (2009). \\
In Fig.~\ref{sfr}, we show the star formation histories of a set of 9 MW-like 
galaxies drawn from our sample. 
In Fig.~\ref{gas}, we show the cumulative stellar and gas masses 
as a function of time for these galaxies, whereas in Fig. ~\ref{infall} we show the time evolution of the
gas accretion rates. \\
The predicted abundance ratios between any chemical element and Fe are sensitive to the parameters 
we have assumed to describe the type Ia SN rate. In particular, having assumed in this case a Salpeter IMF 
constant in time, the most relevant parameter in this study is the type Ia realization probability $A_{Ia}$. 
In the literature, this quantity is not well-constrained, but it is generally considered 
as a free parameter tuned in order to reproduce the local type Ia SN rate (Matteucci et al. 2006; Calura \& Matteucci 2006). 
In Fig.~\ref{snum}, we show the present-time type Ia SN rate, expressed in SNuM\footnote{1 SNuM=$1$ SN $ century ^{-1} 10^{-10}M_{\odot}$}, 
predicted for our selected MW-like galaxies and compared to an observational range for  the Milky-Way 
global Ia SN rate. The observational range is computed by assuming a SN rate of 0.3-0.4 $1/cen$
(van den Bergh \& Tamman 1991) and a present-day disc stellar mass of $\sim 5 \times 10^{10} M_{\odot}$ (Mera et al. 1998). 
This figure shows that the assumption of a Ia SN frequency $A_{Ia}=0.002$ provides 
present-day type Ia SN rates in reasonable agreement with the observations (upper panel of Fig.~\ref{snum}) 
whereas by assuming $A_{Ia}=0.004$ we severely overestimate the observed type Ia SN rate (lower panel of Fig.~\ref{snum}). 
From this moment on, $A_{Ia}=0.002$ 
be our reference value for the type Ia SN realization probability. \\
The chemical abundances obtained for the galaxies selected according to the criteria described at the beginning of this section are now 
compared to the abundances observed in Milky Way stars.

\begin{figure*}
\centering
\vspace{0.001cm}
\epsfig{file=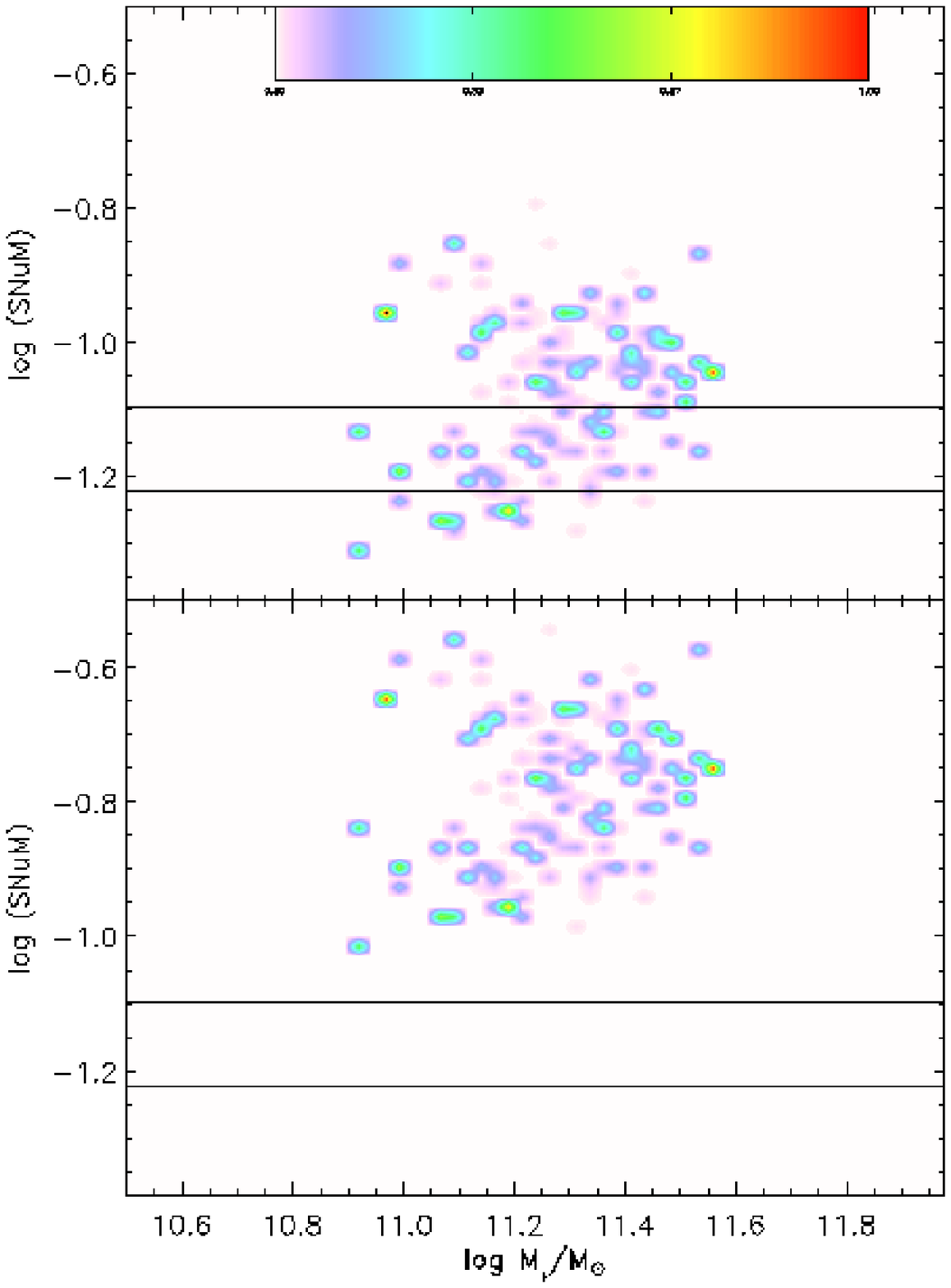,height=10cm,width=10cm}
\caption{Predicted type Ia SN rate, expressed in SNuM (1 SNuM = $1$ SN $ century ^{-1} 10^{-10}M_{\odot}$) 
for selected MW-like galaxies compared to an estimate of the 
type Ia SN rate for the Milky Way (black solid lines). The colour code, shown by the bar at the top of the Figure,  
represents the predicted number of galaxies with a given SNuM and a stellar mass $M_{*}$, 
normalised to the total number of galaxies with  that  stellar mass. In the upper and lower panels, 
the SN rate is computed by assuming $A_{Ia}=0.002$ and $A_{Ia}=0.004$, respectively.
}
\label{snum}
\end{figure*}
\begin{figure*}
\centering
\vspace{0.001cm}
\epsfig{file=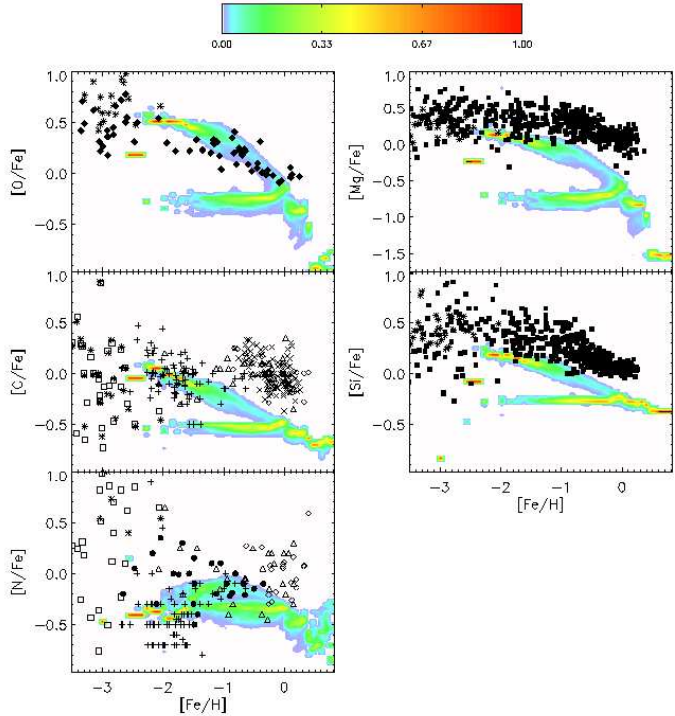,height=10cm,width=10cm}
\caption{Predicted distribution of the abundance ratios [X/Fe] vs [Fe/H] for selected MW-like galaxies 
for several chemical elements compared with 
observations from various authors. 
The colour code, shown by the bar at the top of the Figure, represents the predicted number of stellar populations belonging to MW-like 
galaxies with a given 
abundance ratio at metallicity [Fe/H], 
normalised to the total number of stellar populations at that metallicity. 
\emph{Observational data}: asterisks: Cayrel et al. (2004);
plus signs: Carbon et al. (1987); open squares: Spite et al. (2005), open triangles: Laird (1985a; 1985b), 
the crosses: Tomkin et al. (1995), open diamonds: Clegg et al. (1981), 
solid circles: Israelian et al. (2004), solid diamonds: Fran\c cois et al. (2004), finally the solid squares 
are from a compilation of data by Cescutti (2008). 
}
\label{x_fe}
\end{figure*}
\begin{figure*}
\centering
\vspace{0.001cm}
\epsfig{file=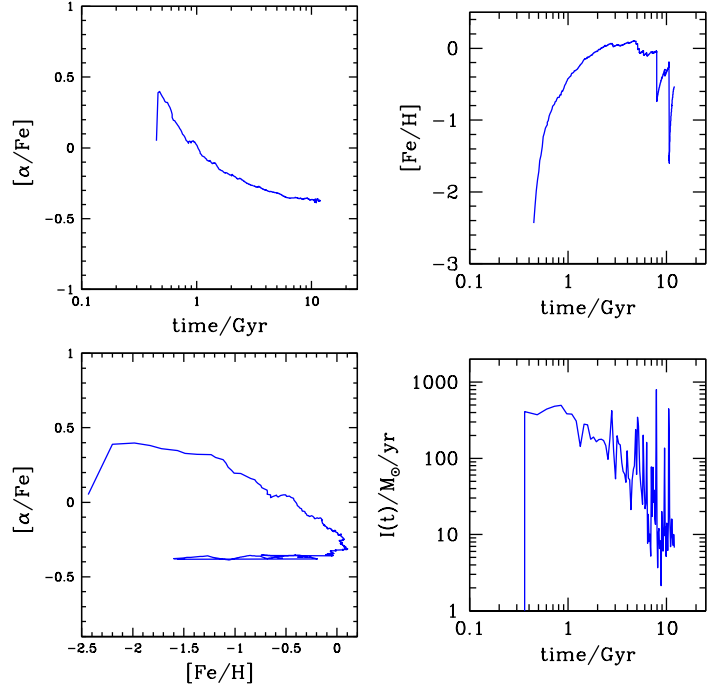,height=10cm,width=10cm}
\caption{From top-left, clockwise: [$\alpha$/Fe] versus time, [Fe/H] versus time, Infall rate versus time and
[$\alpha$/Fe] versus [Fe/H] for one of the selected MW-like galaxies discussed in Section ~\ref{MW}}
\label{afe_exa}
\end{figure*}
\begin{figure*}
\centering
\vspace{0.001cm}
\epsfig{file=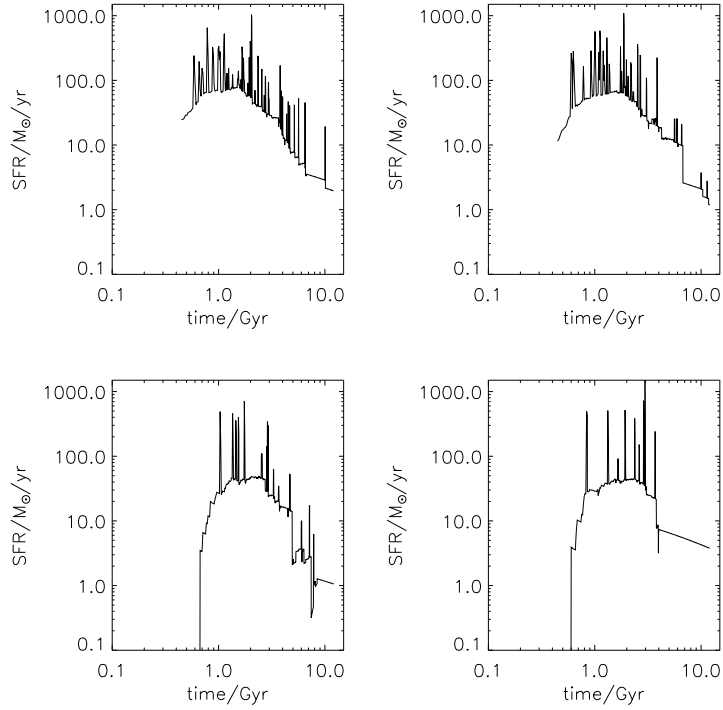,height=10cm,width=10cm}
\caption{ Star Formation histories of MW-like galaxies 
whose mass assembled at $z=2$ is at least $75\%$ of their present mass.}
\label{sfr_noh}
\end{figure*}
\begin{figure*}
\centering
\vspace{0.001cm}
\epsfig{file=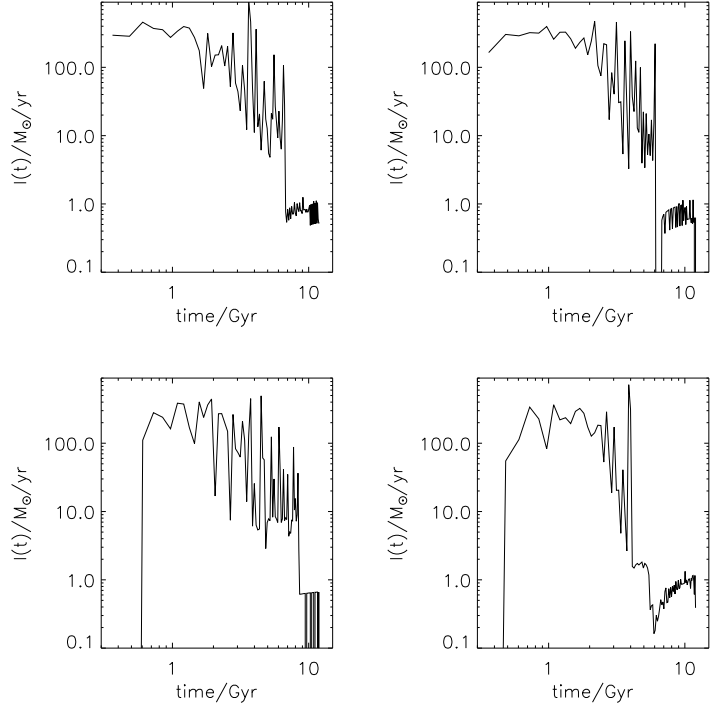,height=10cm,width=10cm}
\caption{Time evolution of the increment rate $I(t)$ for 
MW-like galaxies whose assembled mass at $z=2$ is at least $75\%$ of their present mass. }
\label{inf_noh}
\end{figure*}
\begin{figure*}
\centering
\vspace{0.001cm}
\epsfig{file=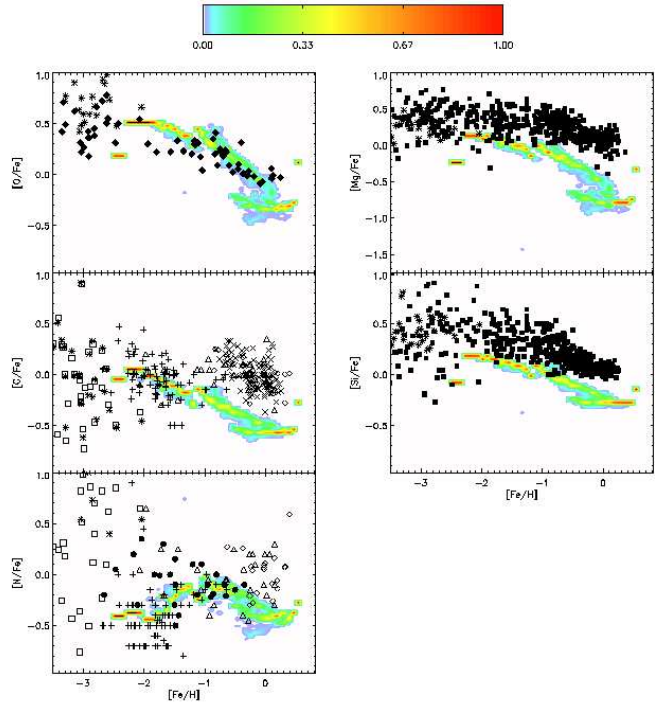,height=10cm,width=10cm}
\caption{Predicted distribution of the abundance ratios [X/Fe] vs [Fe/H] for MW-like galaxies 
whose mass assembled at $z=2$ is at least $75\%$ of their present mass.  
for several chemical elements compared with 
observations from various authors. 
The colour code, shown by the bar at the top of the Figure,  represents the predicted number of stellar populations belonging to 
MW-like galaxies born with a given 
abundance ratio and with metallicity [Fe/H], 
normalised to the total number of stellar populations with that metallicity. Observational data as in Fig.\ref{x_fe}}
\label{x_fe_noh}
\end{figure*}

\begin{figure*}
\centering
\vspace{0.001cm}
\epsfig{file=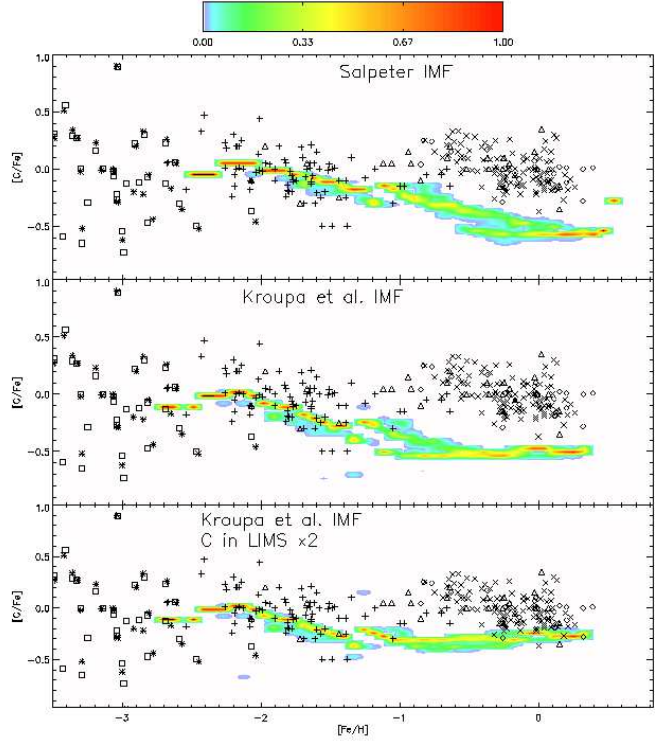,height=10cm,width=10cm}
\caption{Effects of varying the stellar yields and the IMF on the predicted 
[C/Fe] vs [Fe/H] for selected MW-like galaxies.
We show our results obtained assuming a Salpeter IMF and the standard stellar yields described in Sect.\ref{chem} (upper panel), 
a Kroupa et al. (1993) IMF and standard yields (middle panel), a Kroupa et al. (1993) IMF and 
the C yields of low and 
intermediate stars increased by a factor 2. Colour code and observational data as in Fig.\ref{x_fe}.}
\label{cfe_fe}
\end{figure*}
\begin{figure*}
\centering
\vspace{0.001cm}
\epsfig{file=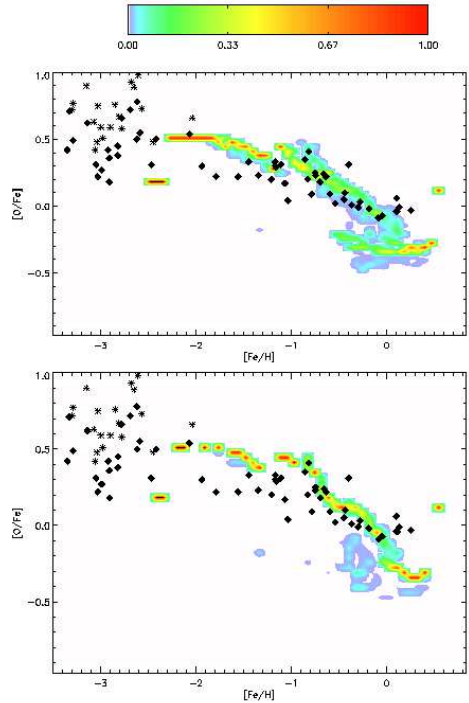,height=10cm,width=10cm}
\caption{Predicted [O/Fe] vs [Fe/H] for Milky Way galaxies selected on the basis of their gas accretion history (upper panel) and 
for the ones presenting a ratio between the stellar mass in the disc and total stellar mass $M_{*,d}/M_{*,tot}=0.7 - 0.8$ (lower panel). 
Colour code and observational data as in Fig.\ref{x_fe}. 
}
\label{ofe_fe_b_t}
\end{figure*}

\begin{figure*}
\centering
\vspace{0.001cm}
\epsfig{file=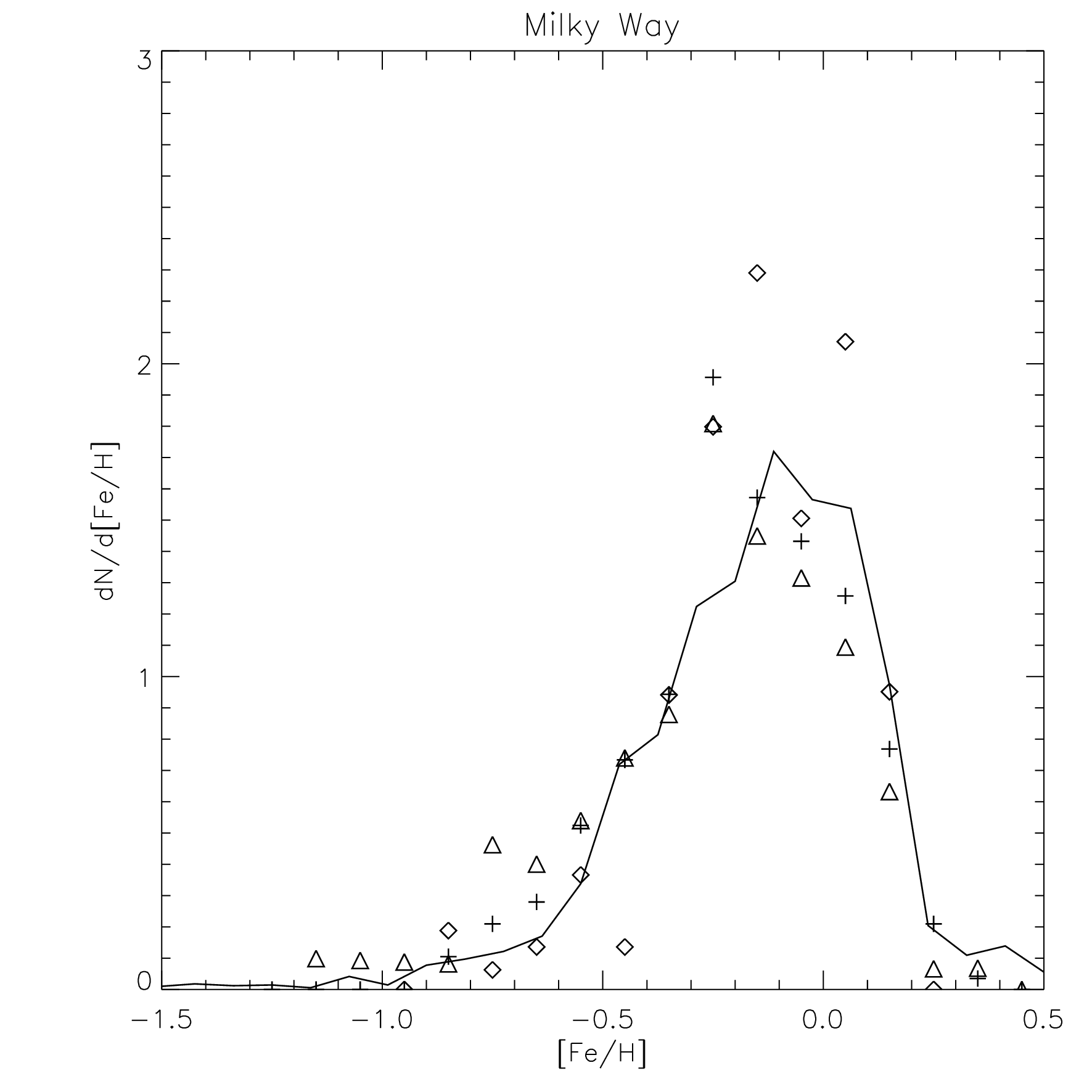,height=10cm,width=10cm}
\caption{Predicted cumulative stellar metallicity distribution 
of selected Milky-Way like galaxies (solid line) compared to local observations from various
authors. The model predictions and the observational data are normalised by imposing that the 
subtended areas have all the same value, equal to unity. Observations are from 
Wyse \& Gilmore (1995) (crosses); Rocha-Pinto \& Maciel (1996) (triangles) and Jorgensen (2000) (diamonds)}
\label{dndfe}
\end{figure*}
In Fig.~\ref{x_fe}, 
we show the predicted abundance ratios vs metallicity for our sample of MW-like galaxies, compared to 
a set of observations of abundance ratios in MW stars for various elements. 
The predictions are represented by the colour code at the top of  Fig.~\ref{x_fe}, 
expressing the predicted number of stellar populations belonging to galaxies with given 
abundance ratio at metallicity [Fe/H], 
normalised to the total number of stellar populations at that metallicity.
The symbols are the observational values from several sources (for further details, see 
caption of Fig.~\ref{x_fe}).
In Fig.~\ref{x_fe}, the observational data are single-star abundances, i.e. each point 
represents the abundance ratios observed in a single local star. On the other hand, in our model 
we cannot resolve single stars, hence we consider the average abundances 
in a stellar population and we compare them to observations. For each selected MW-like 
galaxy, in the time interval $dt$ a stellar mass $dM_*$ will be formed, characterized 
by a given metallicity (i.e. [Fe/H]) and particular abundance ratios. 
In Fig. ~\ref{x_fe}, the coloured regions represent the predicted [X/Fe] - [Fe/H] 
distributions computed considering each stellar population of mass $dM_*$ born in each selected 
MW-like galaxy. This is very similar to what is done in chemical evolution models for the solar 
neighbourhood (see Matteucci 2001), barring the fact that in that case single tracks, representing the 
predictions for one single star formation history, 
are 
plotted in the [X/Fe]-[Fe/H] diagrams. In our case, instead of having one single evolutionary track, 
we have various model tracks, each one drawn from a particular star formation history for a 
MW-like galaxy. All these tracks populate the 
coloured areas of Fig. ~\ref{x_fe} .\\
It is important to note that we cannot explore the metallicity region [Fe/H]$\le -2.5$. 
To study chemical evolution at metallicities lower than this value, 
it would be necessary to consider finite stellar lifetimes also for 
 massive stars. However, the metallicity range 
$-2.5 \le [Fe/H]\le 0$ encompasses the Fe abundances of all thin and thick disc stars, hence it is useful 
to investigate the chemical evolution of MW-like galaxies during their major disc phases. 
In Fig.~\ref{x_fe}, in the [O/Fe] vs [Fe/H] plot 
the predictions indicate the presence of 
two main populations of stars. 
The first population has the [O/Fe] ratio 
in anti-correlation with the [Fe/H] and overlaps 
with the observation of local stars. 
Beside this, Fig.~\ref{x_fe} shows the presence of 
a horizontal population of stars with metallicity $-2. \le [Fe/H] \le -0.2$ and 
constant [O/Fe]$\sim -0.3$.  A similar behaviour  is visible 
also in the [C/Fe]$-$[Fe/H], [Mg/Fe]$-$[Fe/H], and [Si/Fe]$-$[Fe/H] plots. 
This is due to substantial 
late increment episodes, very frequent in selected MW-like galaxies 
(see Fig.~\ref{infall}). The late accretion  of pristine gas 
has the effect of decreasing the 
metallicity [Fe/H] of each stellar population born immediately after the infall event, having 
little effect on the [O/Fe] ratio, since both O and Fe are diluted by the same amount. 
Note that none of the observational data in 
Fig.~\ref{x_fe} shows this peculiar behaviour.   
To better  understand this point, 
in Fig. ~\ref{afe_exa} we show for one of the selected MW-like galaxies  
the time evolution of the average 
interstellar 
[$\alpha$/Fe] and [Fe/H], of the total infall rate and the [$\alpha$/Fe] vs [Fe/H] diagram. 
From the [Fe/H] vs time  and infall vs time plots, one can clearly see that 
in correspondence of two major gas accretion episodes at 7.5 Gyr and $\sim$ 11 Gyr, 
the [Fe/H] drops abruptly whereas the [$\alpha$/Fe] remains unchanged. 
In the [$\alpha$/Fe] - [Fe/H] diagram, the late accretion episodes and the consequent decrease of interstellar [Fe/H]
are visible in the 'turn-off ' of the track at [Fe/H] $\sim$ 0, 
where the curve bends back towards lower [Fe/H], at constant [$\alpha$/Fe] $\sim$ -0.5. 
The decrease of [Fe/H] at constant  [$\alpha$/Fe] stops at [Fe/H] $\sim$ -1.7, where the track bends forward, moving towards  higher 
[Fe/H] values. \\
To investigate this point, we 
perform a further selection on the sample 
of MW-like galaxies. 
The basic reason of the presence of the horizontal population of stars in the 
[O/Fe]-[Fe/H] is late accretion episodes. 
Now we focus on a given redshift, in order to understand which fraction of the present-day 
mass must be  assembled in galaxies presenting chemical features the most similar to those  of the MW  as possible. 
This test may be useful to give constraints  on the gas accretion history of the MW galaxy. 
We may focus on $z=2$, which may be used a lower limit to the highest redshift of major 
merging for MW-like galaxies (see e.g. Colavitti et al. 2008). 
It is necessary that at least 75 \% of the present-day mass must be assembled at $z=2$, 
 in order to have the horizontal population vanished.  
In Fig. \ref{sfr_noh}, we show the SF histories of these galaxies selected on the basis of their accretion history.  
The accretion histories of these galaxies are shown in Fig. \ref{inf_noh}. 
 Basically, these galaxies do not experience major 
accretion episodes at epochs after $\sim 6$ Gyr, at variance with 
most of the systems represented in Fig. ~\ref{infall}, whereas their star formation histories 
show low star formation rates at late times. 
In Fig.\ref{x_fe_noh}, we show the the predicted abundance ratios vs metallicity for all these galaxies. 
From Fig.~\ref{x_fe_noh}, we see that in most of the cases, the observed behaviour of the 
abundance ratios measured in Milky Way stars is satisfactorily reproduced. \\
Our results suggest that the abundance ratios observed in the Milky Way stars exclude that 
our Galaxy experienced major pristine gas accretion episodes within the last few Gyrs, either due to merging with 
gas-rich, low-metallicity dwarf galaxies, or due to infall of pristine gas. \\
Our results are in agreement with a large set of previous papers, 
indicating that the merging history of the MW must be completed at early times and that 
no significant infall episode took place in the last few Gyrs, 
most of them based on the theoretical 
interpretation of abundances observed in MW disc and halo 
stellar populations or present-day properties of the MW galaxy (Prantzos \& Silk 1998, Boissier \& Prantzos 1999;  
Hernandez et al. 2001; Alibes et al. 2001; Chiappini et al. 2001; Naab \& Ostriker 2006; De Lucia \& Helmi 2008; 
Colavitti et al. 2008)  and on kinematical arguments (e.g. Wyse 2009; Gilmore et al. 2002). \\
In Fig.\ref{x_fe_noh}, 
the fact that for a few elements the predictions are shifted 
downwards with respect to the observed abundance ratios is mainly due to uncertainties 
related to the stellar yields used here. 
The [Mg/Fe]
decreases by about 1.5 dex instead of less than 0.5 dex from
[Fe/H]=-2 to [Fe/H]$\sim$ 0. 
 Concerning the elements Mg and Si, it is well known from chemical evolution studies  
that the yields of Woosley \& Weaver (1995) used 
here tend to underestimate the [Mg/Fe]-[Fe/H] and [Si/Fe]-[Fe/H] relations.
Franc\c ois et al. (2004) reproduced the local 
observed abundance pattern by means of a model for the solar neighbourhood, increasing the yields of Mg by a factor of 7 and 
the yields of Si by a factor of 2, only for stars with masses $m>40 M_{\odot}$ (see also Timmes, Woosley \& Weaver 1995). 
Also the adopted C yields do not appear to be sufficient for reproducing the zero-point of the observed 
[C/Fe] - [Fe/H] relation and we underestimate the [C/Fe] on the whole metallicity range. 
Furthermore,  The observed [C/Fe] exhibits a
flat behaviour throughout the whole [Fe/H] range,  whereas the SAM 
results show a remarkable anti-correlation. 
This may be certainly due to the adopted stellar yields, but 
also the IMF is likely to play some role in  the predicted [X/Fe] - [Fe/H] relations. 
In Fig.~\ref{cfe_fe}, we show how different assumptions regarding the IMF and the stellar yields may affect 
our predictions in the [C/Fe] - [Fe/H] plot.
With a Salpeter IMF and standard yields, our match to the data is very poor. 
The use of a Kroupa et al (1993) IMF causes a flattening of the predicted [C/Fe] - [Fe/H] at metallicities [Fe/H]$>-1$. 
This is due to the fact that with the Kroupa IMF the relative fraction of  intermediate mass stars, 
i.e. stars with masses 
$2\le M/ M_{\odot} \le 8$ , representing the most important C producers, is larger than with the Salpeter. 
If we assume a Kroupa et al. (1993) and we artificially 
increase the yields of intermediate  mass stars by a factor 2, 
all the predictions at metallicity [Fe/H]$>-1$ shift upwards and overall 
the predicted [C/Fe] - [Fe/H] becomes flatter, in better agreement with the observations. }\\
Romano et al. (2005) showed in detail the relative roles of the stellar yields and of the IMF in determining the abundance ratios, 
suggesting that the Salpeter IMF is not appropriate to reproduce most local chemical evolution constraints. 
However, a  detailed investigation of the stellar yields and of the stellar IMF requires a fine-tuning of these parameters 
and is beyond the aims of the present paper, 
which is instead focused on the chemical evolution of galaxies within a cosmological framework. \\
We have verified that the use of a morphological criterion to select 
MW-like galaxies has a minor impact on our results. 
Among our sample of galaxies matching the chemical properties of MW-like galaxies, 
we can calculate the ratio between the stellar mass in the disc and the total stellar mass $M_*$ 
in the following way. 
The neutral gas mass of the MW disc is $\sim 1\, \cdot 10^{10} M_{\odot}$ 
(Prantzos \& Silk 1998; Carroll \& Ostlie 1996), whereas the stellar mass is $\sim 5 \, \cdot 10^{10} M_{\odot}$, with a 
ratio between the gas mass and the stellar mass $\sim 0.2$.  
We use this value to estimate the stellar mass in the disc of our selected MW-like galaxies, which 
is, since  all the cold gas is in the disc, 
$M_{*,d} = M_{g}/0.2$. By assuming that the stellar mass of the Milky Way Bulge is $1 - 2 \cdot 10^{10} M_{\odot}$ 
(Ballero et al. 2008, Reshetnikov 2000),  the observed disc-to-total stellar mass ratio is 
$M_{*,d}/M_{*,tot}=   0.7 - 0.8$. 
Of the MW-like galaxies selcted on the basis of their gas accretion history, we now select those with 
$M_{*,d}/M_{*,tot}$ values $\sim 0.7 - 0.8$. In Fig.~\ref{ofe_fe_b_t}, we show the predicted [O/Fe]-[Fe/H] relation for 
all the MW-galaxies selected on the basis of their accretion history (upper panel) and 
on the basis of the $M_{*,d}/M_{*,tot}$ ratio (lower panel), and he results are nearly the same. 
This result is robust against  any morphological selection criterion we use to select MW-like galaxies. \\
In Fig.~\ref{dndfe}, we show the predicted cumulative 
stellar metallicity distribution (SMD hereinafter) for our 
MW-like galaxy sample, compared to observations of solar neighbourhood stars 
from various authors. The predicted SMD has been computed by taking into account the star formation 
histories of the galaxies whose assembled mass at $z \sim 2$ is greater than $75 \%$ of their present mass. 
The predicted SMDs  and the observations are normalised by requiring that the areas subtended by each curve 
have all the same values, equal to unity. 
The agreement between the observed and predicted SMD is remarkable. 
The position of the peak is 
much  sensitive to the type Ia
SN normalization. With our assumption of $A_{Ia}=0.002$, 
the position of the peak of the SMD is satisfactorily reproduced.
We also reproduce accurately the low-metallicity tail of the SMD, i.e. we predict the 
correct fraction of stars in the metallicity range [Fe/H]$ \le -0.7$. 
The high metallicity tail, i.e. the number o stars with [Fe/H]$ \ge 0.2$, is in reasonable 
agreement with the observations. 
It is worth to note that the assumption of the IMF may have some effect on the predicted SMD. 
Fe is produced mainly by Type Ia SNe, whose rate depends on the 
the product of the parameters $k_{\alpha}$ and $A_{Ia}$ of eq.~\ref{SNRIa}.  
$k_{\alpha}$ depends on the IMF, hence a change of the IMF would require a re-tuning of the parameter 
$A_{Ia}$ in order to reproduce the correct present-day SN rate. 
However, it seems reasonable to assume 
that the quantity $k_{\alpha} \cdot A_{Ia}$ 
is  constant and fixed 
in order to reproduce 
the present type Ia SNR in MW-like galaxies. In this case, the IMF assumption is not relevant.\\ 
It may be interesting to compare our results with the ones obtained previously by other authors. 
A previous relevant theoretical study of the chemical evolution of spiral 
 galaxies in a cosmological framework is the one of Nagashima \& Okamoto (2006). In this paper, 
the authors find a good agreement between the predicted [O/Fe] vs [Fe/H] 
obtained for a sample of MW-like spiral galaxies and the observations of the abundances in 
local disc stars. These authors obtained also a good match between the predicted and the observed 
SMD. However, to model the type Ia SN rate and Fe production, Nagashima \& Okamoto (2006) 
 considered a constant value for the delay time of the explosion, 
which seems rather unrealistic, as 
indicated by a large amount of observational and theoretical studies of the type Ia SN rate in local 
and distant galaxies (Mannucci et al. 2006; Matteucci et al. 2006; Sullivan et al.2006; 
Valiante et al. 2009) which, on the other 
hand, point towards continuous delay-time distributions for type Ia SN explosions. \\
Other interesting results are those obtained by Colavitti et al. (2008). In this paper, 
the authors start from cosmological numerical cold dark matter 
simulations and, from the merging history of the parent halos, these authors infer the 
infall history of spiral galaxies, which is implemented in a numerical chemical evolution code and 
whose evolution is followed a-posteriori. Colavitti et al. (2008) find an overall good 
agreement between their predicted abundances and the observational constraints. However, 
the merging histories are determined by means of simulations with a low time resolution. 
Furthermore, the assumption of a linear scaling between the baryonic accretion history and the 
cold-dark matter merging history may represent a raw approximation, as well as 
the fact that pure dark matter simulations 
permit to explore spatial scales of the order of $\sim 1 Mpc$, considerably higher than the scales 
of a few kpc studied in the paper by Colavitti et al. (2008). \\
The one described in this section is the first attempt to reproduce the abundance
ratios for such a wide set of chemical elements 
by means of a hierarchical SAM 
taking into account the lifetimes of low and intermediate mass stars and type Ia SNe. 
Overall, from the various results discussed in this section, we find that most of the 
chemical evolution constraints considered 
here and concerning the Milky-Way galaxy are satisfactorily reproduced. 
We find that the results discussed in this section are very encouraging. 
The agreement between our predictions and the observations 
is remarkable, in particular concerning the [O/Fe]-[Fe/H] diagram and the SMD, 
given the very little fine tuning of the parameters performed so far. 
For other elements shown in Fig.~\ref{x_fe}, 
better fits to the data could be achievable by means of a fine-tuning of the adopted stellar yields, as is performed  sometimes 
in theoretical studies of the chemical evolution of the Solar Neighbourhood (e.g. Fran\c cois et al. 2004).

\begin{figure*}
\vspace{0.001cm}
\epsfig{file=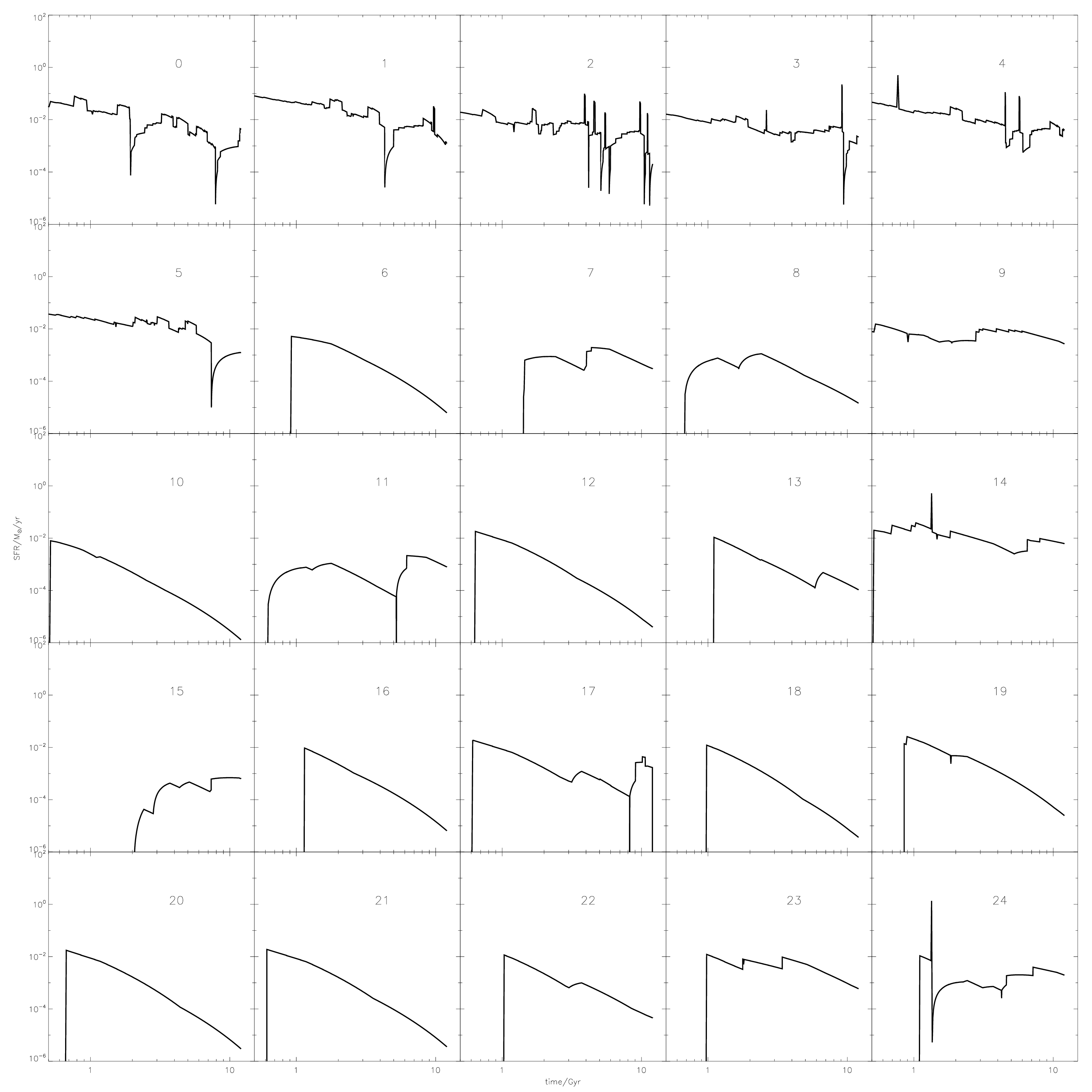,height=15cm,width=15cm}
\caption{ Star Formation histories of a few dwarf  galaxies selected according the criteria described in 
Sect.~\ref{dsph}.}
\label{sfr_dsph}
\end{figure*}
\begin{figure*}
\vspace{0.001cm}
\epsfig{file=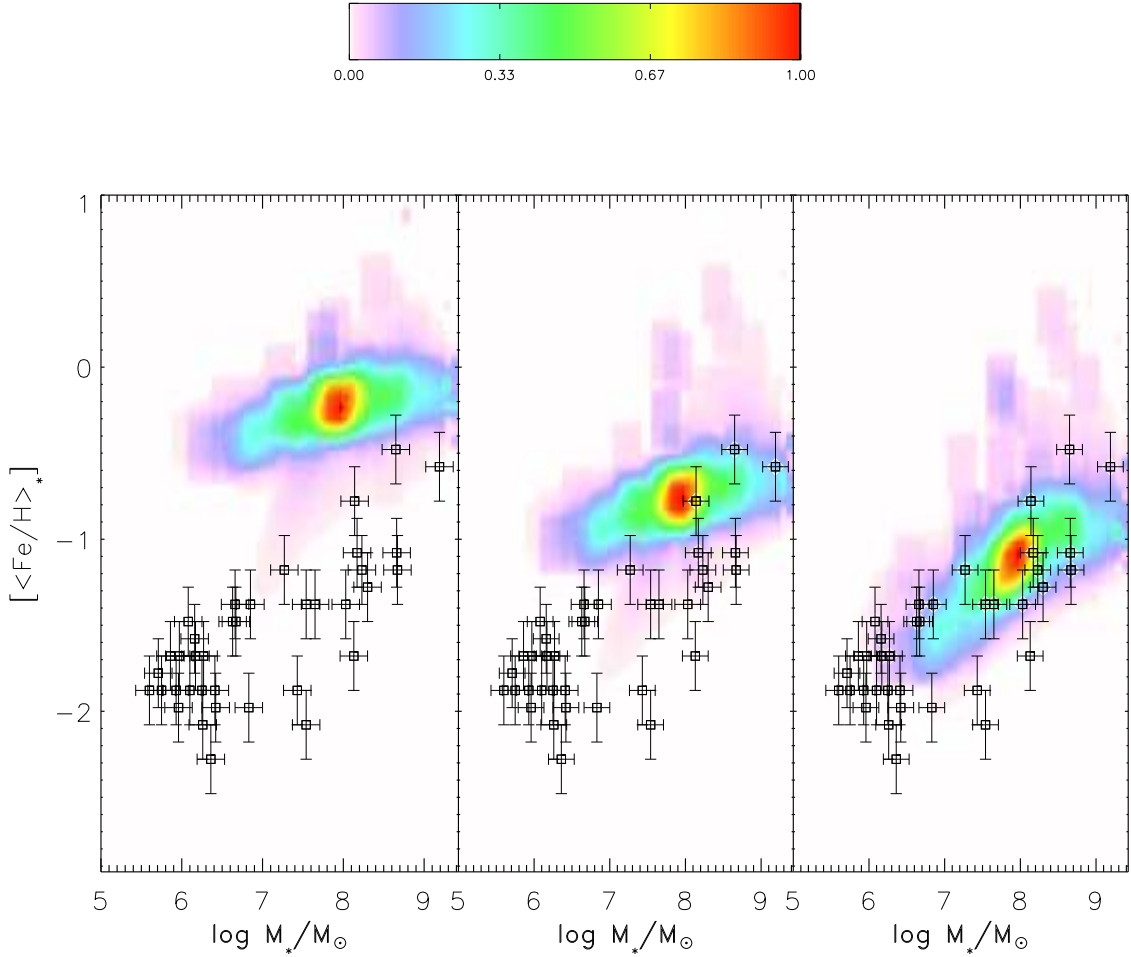,height=15cm,width=15cm}
\caption{Predicted distribution of the average present-day 
stellar metallicity in dwarf galaxies versus present stellar mass.  
The colour code, shown by the bar at the top of the Figure, represents the predicted number of galaxies with given 
$[<Fe/H>_{*}]$ with stellar mass $M_{\odot}$. 
The open squares with error bars are the observations in local dSphs, taken from Woo et al. (1998). 
In the left panel, 
we assume that the chemical composition of the ouflows $X_{i,out}$ is the same as the one of ISM and $A_{Ia}=0.002$  
as assumed for MW-like galaxies. 
In the central panel, we assume chemical compositions of the outflows as above and $A_{Ia}=0.0002$. 
In the right panel, we assume $X_{i,out} = \xi X_{i}$, with  $\xi=10$, and  $A_{Ia}=0.0002$. 
}
\label{fesigma}
\end{figure*}

\begin{figure*}
\vspace{0.001cm}
\epsfig{file=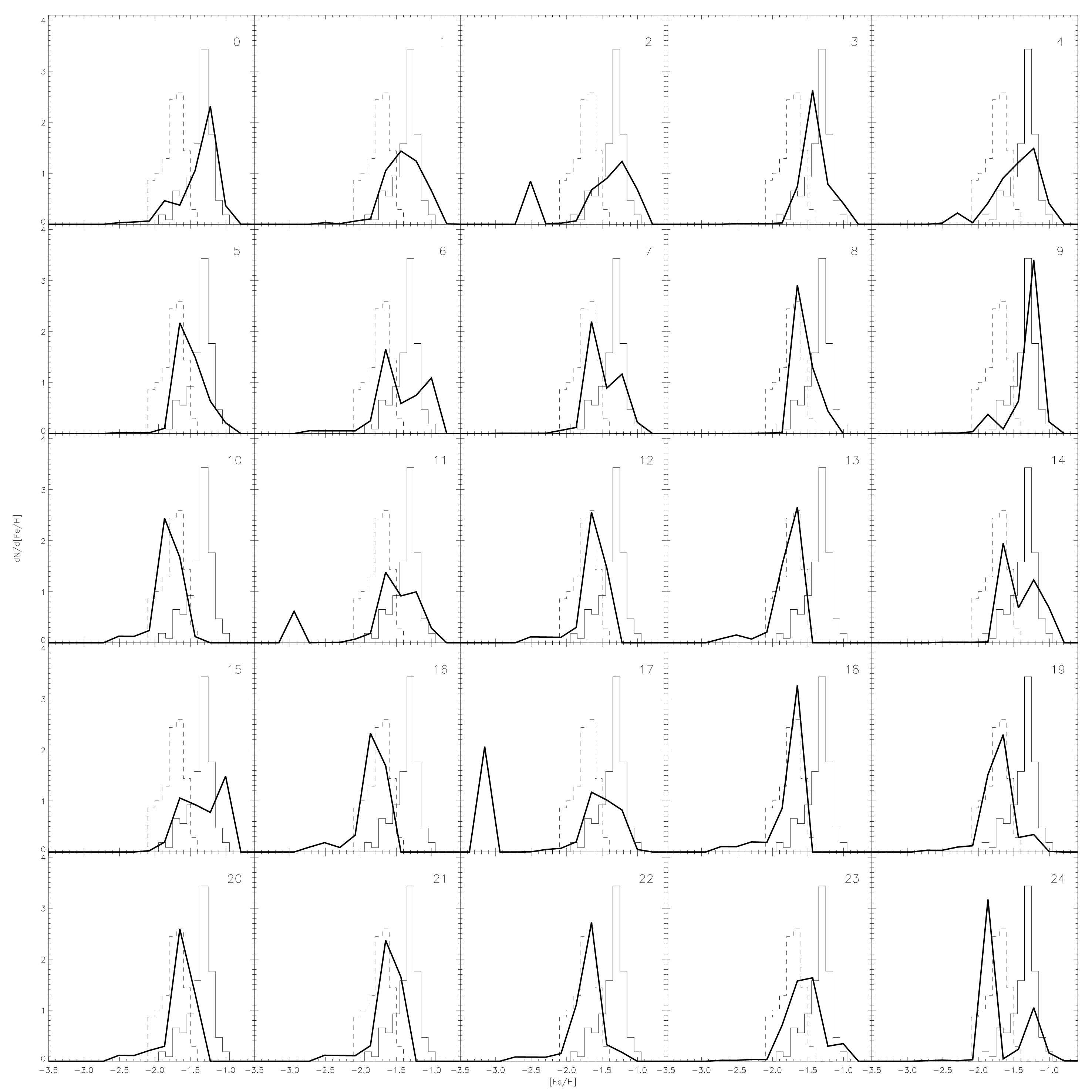,height=15cm,width=15cm}
\caption{Stellar Metallicity distributions computed for the same set of dwarf galaxies presented 
in Fig.~\ref{sfr_dsph} (solid thick lines), compared with observational SMDs obtained for the Draco dSph  
(solid thin histograms) and for the Ursa Minor dSph (dashed thin histograms, see Bellazzini et al. 2002).}
\label{dndfe_dsph_m}
\end{figure*}
\begin{figure*}
\centering
\vspace{0.001cm}
\epsfig{file=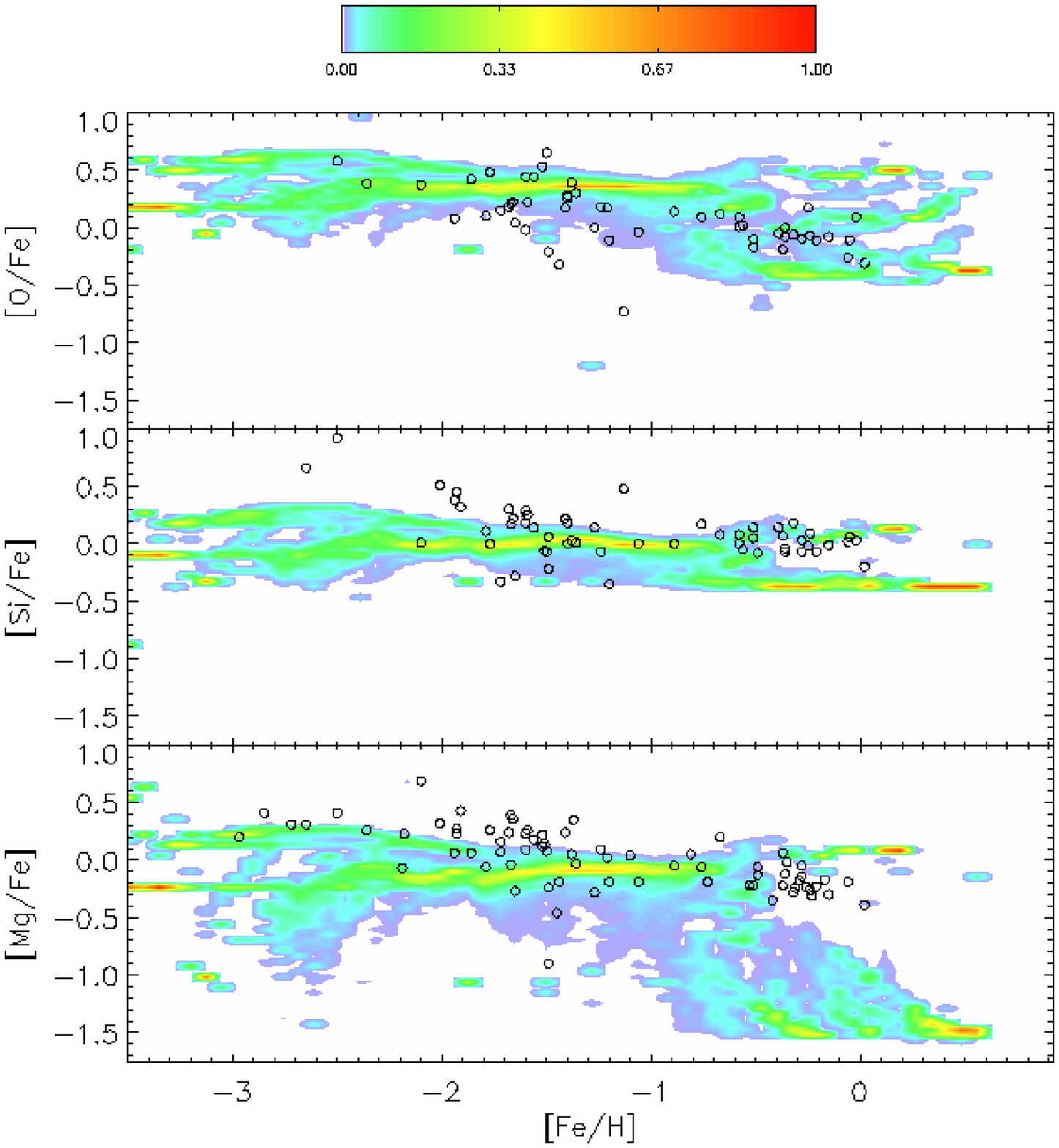,height=10cm,width=10cm}
\caption{Predicted distribution of the abundance ratios [X/Fe] vs [Fe/H] 
for dwarf spheroidal galaxies. 
 The colour code, shown by the bar at the top of the Figure,  represents the predicted number of stellar populations belonging to dwarf galaxies 
with a given 
abundance ratio at metallicity [Fe/H], 
normalised to the total number of  stellar populations  with that metallicity.
The 
Observational data (open circles) are observations in local dwarf spheroidals 
taken from a compilation by Lanfranchi \& Matteucci (2003). 
}
\label{x_fe_dsph}
\end{figure*}

\subsection{Abundances in dwarf galaxies}
\label{dsph}
\begin{figure*}
\centering
\vspace{0.001cm}
\epsfig{file=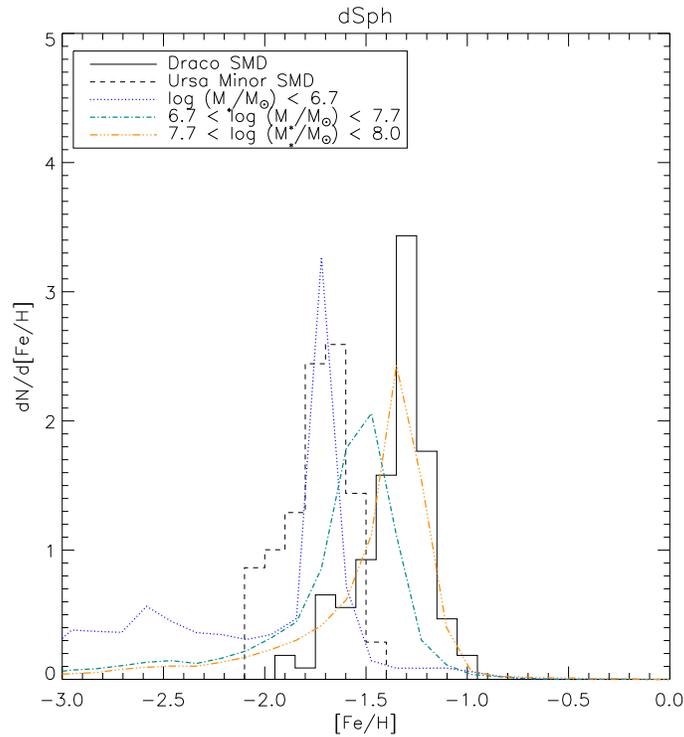,height=10cm,width=10cm}
\caption{Predicted cumulative stellar metallicity distribution 
of selected dwarf galaxies belonging to three different mass bins (solid thick lines) compared 
with observational SMDs obtained for the Draco dSph 
(solid histogram) and for the Ursa Minor dSph (dashed histograms).
}
\label{dndfe_dsph}
\end{figure*}

In this section, we present our results concerning the chemical evolution of dwarf galaxies. 
In particular, we focus  on the chemical abundances in dwarf spheroidal (dSph) galaxies, since for this kind of objects 
a large set of observational data is available from observations 
of the dSph in the Local Group. 
In this case, 
the selection on our galaxy catalogue is performed on the basis of the present stellar mass $M_{*}$, 
considering all the systems with $10^{6} \le M_{*} /M_{\odot} \le 5 \cdot 10^{9}$, broadly  
corresponding to the stellar mass range of local dwarf galaxies (Mateo 1998). 
In Fig.~\ref{sfr_dsph}, we see the time evolution of the SFR for a few selected examples. 
One striking feature of Fig.~\ref{sfr_dsph} is the variety of the star formation histories 
of the selected galaxies. 
For some systems (e.g. the objects labelled 18 and 20 in Fig.~\ref{sfr_dsph}), 
the SFH is maximum at the beginning, then smothly and constantly decreasing. 
Some others have more than one peak (e.g. 2, 11), 
mimicking the starbursty behaviour of some real dSph galaxies, whose complex 
star formation history is usually inferred from the observed colour-magnitude diagrams 
(van den Bergh 1994; Hernandez, Gilmore, Valls-Gabaud 2000; Dolphin et al. 2005). \\
An important plot  is the stellar metallicity vs stellar mass  (Fig.~\ref{fesigma}), 
allowing us to test the star formation histories of our dSphs and also to tune some fundamental 
parameters regarding their chemical evolution. 
In this figure, the stellar metallicity is represented by 
the average stellar Fe abundance $[<Fe/H>_*]$. 
This plot is of great interest since it is 
analogous to a mass-metallicity plot, 
with the difference that in general, in the mass-metallicity plots 
the stellar mass and the interstellar metallicities 
are represented, the latter usually derived  by means of observations of O emission lines in brilliant 
$H_{II}$ regions of star-forming galaxies (Maiolino et al. 2008). 
Local dSphs are weakly star-forming systems and do 
not allow observations of emission lines in $H_{II}$ regions, 
however  a plot like such of Fig.~\ref{fesigma} provides us with similar indications as a  $O/H$ vs $M_{*}$ plot, 
since $[<Fe/H>_{*}]$ is approximately :
\begin{equation}
<Fe/H>_{*}\simeq \frac{\int (Fe/H)(t) dM_{*}}{\int dM_{*} }=\frac{\int (Fe/H)(t) \psi(t)dt}{M_{*,tot} }
\end{equation}
i.e. the integral of the interstellar Fe abundance over 
the star formation history (see Thomas et al. 1999). 
The predicted $[<Fe/H>_*]$-$M_{*}$ relation is compared to the one observed in local dwarf galaxies by Woo et al. (2008). 
The observations indicate that the local dwarf galaxies clearly follow a mass-metallicity relation, i.e. that the 
smallest galaxies have the lowest metallicities, and that the stellar metallicity increases with stellar mass. 
In the left panel of Fig.~\ref{fesigma}, we have assumed that dwarf galaxies can undergo mass loss under the same conditions as 
spiral galaxies, i.e. the chemical composition of the outflow is the same as the one of the ISM, 
$O_{i} (t) = X_{i} (t) \cdot O(t)$, and for the type Ia SN realization probability the same value 
as assumed for spirals, $A_{Ia}=0.002$. 
In most of the cases, the model dwarf galaxies are considerably more metal-rich than the observed ones. 
Furthermore, the predicted mass-metallicity relation is much flatter than the observational one. 
In the middle panel of Fig.~\ref{fesigma}, we show the predicted $[<Fe/H>_*]$-$M_{*}$ 
relation computed by assuming a chemical composition of the outflows as above and 
a very low value for the type Ia SN realization probability, $A_{Ia}=0.0002$, 
i.e. one tenth of the value used for MW-like galaxies. 
This assumption certainly lowers the zero-point of the predicted mass-metallicity relation, 
however does not affect its shape, which is still flatter than the observed one.   
Both the slope and the zero-point of the predicted mass-metallicity relation can be further improved 
by modifying our assumptions on metal loss in dwarf galaxies. 
In the right  panel of Fig.~\ref{fesigma}, we have assumed preferential loss of metals in the outflows, 
i.e. that for any chemical element different than H and He, the outflow rate in eq.~\ref{chemeq} is given by 
$\xi X_{i} (t) \cdot O(t)$, with $\xi=10$, with  $A_{Ia}=0.0002$. 
As can be seen, this assumption improves considerably the match between the predicted and 
observed mass-metallicity relation for dwarf galaxies. 
One first conclusion on chemical evolution of dwarf galaxies is that, in order to reproduce the 
slope of the observed mass-metallicity 
relation, 
one needs to assume that in dwarf galaxies 
mass loss must occur through 
metal-enhanced outflows. 
The importance of mass loss in determining 
the mass-metallicity relation in dwarf galaxies has already been ascertained (Dekel \& Silk 1986). 
The fundamental role played by metal-enhanced outflows has already been discussed 
by some previous investigations 
on the mass-metallicity and luminosity-metallicity relations in dwarf galaxies by means of semi-analytic galaxy formation models 
(Somerville \& Primack 1999). 
All the following results presented in the remainder of this section 
are computed by assuming metal-enhanced outflows 
in dwarf galaxies. 
A more detailed discussion on the origin and importance of metal enhanced outflows in dwarf galaxies 
will be presented in Sect.~\ref{concl}. \\
It may be interesting to see the effects of the star formation histories shown in Fig. ~\ref{sfr_dsph} 
on the metallicities of the stellar populations of dwarf galaxies. 
In the remainder of this section, we will compare our predictions with observational 
data regarding local dwarf spheroidal galaxies, whose stellar masses are in general of the order of 
$10^8 M_{\odot}$ or lower (Mateo 1998). For this reason, from this point on, in our analysis we will consider only the model galaxies with present 
stellar masses $M_* \le 10^8 M_{\odot}$. 
In Fig.~\ref{dndfe_dsph_m}, 
we can see the predicted SMDs for the 
same objects as  those presented in Fig. ~\ref{sfr_dsph} . 
In this figure, for comparison we plot also the 
SMDs observed in two local dwarf spheroidals, i.e. Draco and Ursa Minor.
The predicted SMDs have been computed with a number of bins comparable with the ones of the data of Bellazzini et al. 2002, 
which have 12-20 bins. 
For  most of the model galaxies  presented in Fig.~\ref{dndfe_dsph_m}, 
one dominant peak is recognizable, which corresponds to a dominant metallicity of the stellar populations of the dwarf galaxies.  
Some objects present two unresolved peaks (e.g. 6, 14), whereas 
others have two clearly distinct peaks at different metallicity values (11, 17). 
All the model galaxies  with a single-peaked SMD or with two very close, non-resolved peaks have in general smooth star 
formation histories, or presenting small oscillations, or a few very short 
prominent peaks in their SFH. 
All the systems presenting two resolved peaks in the SMD have in general bursty SFHs, charactherized  by quite broad and distinct 
peaks (e.g., 11, 17), or frequent short and high peaks overimposed to 
an overall continuous SFH (2). 
It is worth to note that in most of the cases, the positions of the 
peaks of the predicted SMDs agree with the ones of the observed SMDs. 
It is also worth to note that the lowest metallicities which can be probed by our analysis depend on the SF history. 
In Sect. 3.1, we have seen that in MW-like galaxies, we could not explore the metallicity region [Fe/H]$<-2.5$. 
MW-like galaxies have strong SF histories, hence a metallicity [Fe/H]$\sim -2.5$ can be reached in a 
time shorter than the timestep used in our model. On the other hand, for dwarf galaxies, characterised by 
lower SF rates, the lowest metallicity probed is [Fe/H]$<-3.5$, reached 
sometimes on timescales higher than our timestep. Another reason for the low metallicities is the adopted value 
for $A_{Ia}$, which in the case of dwarfs is one tenth of the value assumed for MW-like galaxies.\\
Also in this case, owing to the large set of chemical elements studied here, we 
are able to produce predictions for the abundance ratios of several chemical elements. 
In Fig.~\ref{x_fe_dsph}, we show our predictions for the [O/Fe] vs [Fe/H], [Si/Fe] vs [Fe/H] 
and [Mg/Fe] vs [Fe/H]. 
For these elements, the abundances are  measurable 
in single stars of local dSphs. 
The  predictions represent the abundance ratios of the stellar populations belonging to the selected galaxies, 
as described in Sect. ~\ref{MW}.
From Fig.~\ref{x_fe_dsph}, we can see how, 
for any element considered here, the characteristic 
decreasing trend of [$\alpha$/Fe] vs [Fe/H] is successfully reproduced. 
In our models, this trend is due to the delay between type II SNe explosions and type Ia 
SN explosions, as in classical chemical evolution models taking into account enrichment from type Ia and type II SNe
(Matteucci 2001 and references therein).   \\
While in the [O/Fe] vs [Fe/H] plot the observed data overlap with our predictions, in the other two plots 
the measured [$\alpha$/Fe] are often underestimated, in particular for [Fe/H]$> -1$. 
The same discrepancy was found also when discussing the results for MW-like galaxies, and is due to the adopted stellar yields 
of Woosley and Weaver (1995), which lead the models to  underestimate the [Mg/Fe] and [Si/Fe] also in the solar neighbourhood 
(Fran\c cois et al. 2004). In these plots, a fine-tuning of the stellar yields as a function of the initial stellar 
mass would be required to reproduce the observations.  
Furthermore, we see that our models for dwarf galaxies predict a wider [Fe/H] range than the one spanned by the observational data, 
but this 
plot is not useful to understand the relative fractions of the stellar populations of various metallicities. 
More useful in this regard is Fig.~\ref{dndfe_dsph}, where we show the predicted 
cumulative SMD of all the dwarf galaxies, divided in 3 stellar mass bins (see legend of Fig. ~\ref{dndfe_dsph}), 
compared to the observational SMDs of the Draco and Ursa Minor dSph. 
The predicted SMDs are 
computed by taking into account 
all the stellar populations of any selected galaxy falling in each mass bin. 
The cumulative stellar metallicity distributions computed for various mass bins are in very good agreement 
with the observed ones, at least concerning the positions of the peaks. 
In each mass bin, the predicted SMDs span very broad metallicity ranges. In particular, 
in this case we predict the existence of stars at very low metallicities, [Fe/H]$ \le -2.$.  
By integrating the quantity $dN/d[Fe/H]$ over different metallicity ranges,  it is possible to 
estimate the relative fractions of the stellar populations at various metallicities. 
For the objects in the lowest mass bin, no stellar population presents  [Fe/H]$> -0.7$. 
For the objects in the intermediate and highest mass bins, 
a very small fraction lower than 1 \% have stellar populations 
with [Fe/H]$>0$. \\
Finally, it is interesting to note that we predict the substantial presence of very low metallicity stars with [Fe/H]$< -2.$, 
with percentages of $\sim 40 \%$, $\sim 15 \%$ and $\sim 10 \%$ in the lowest, intermediate and highest mass bin, respectively. \\
Relevant previous studies of the chemical evolution of dwarf spheroidals 
are those of Lanfranchi et al. (2006). Even if not carried by means of an 
ab-initio galaxy formation model, but by means of galactic 
chemical evolution models whose star formation histories are tuned from 
constraints of CMD observations (see Calura, Lanfranchi \& Matteucci 2008), these works are useful since they allow to have some constraints 
on the parameters affecting this kind of analysis. 
In particular, Lanfranchi \& Matteucci (2007) present  
the effects of the wind efficiency on the predicted SMDs of dSphs, 
showing how a model where the galactic wind is suppressed 
leads to too metal-rich stellar populations. 
The results of 
Lanfranchi \& Matteucci (2004; 2007) suggest that a strong wind efficiency is 
necessary to reproduce the abundance ratios and the stellar metallicity distributions of dwarf galaxies. \\
A previous  study 
of the chemical evolution of dwarf galaxies within a cosmological framework is the one of 
Salvadori et al. (2007). These authors model chemical evolution of dwarf galaxies by taking into 
account the stellar lifetimes of low and intermediate mass stars and of massive stars, 
however they do not consider type Ia SN explosions. Salvadori et al. (2007) explain 
the observed decline of [$\alpha$/Fe] vs [Fe/H] in local dSphs by means 
of differential galactic winds. This result is apparently in contrast with a large variety of 
chemical evolution studies, which underline the importance of the time delay of type Ia SN 
explosions in explaining the decrease of the  [$\alpha$/Fe] vs [Fe/H] in local dSphs 
(Recchi et al. 2001; Ikuta \& Arimoto 2002; 
Lanfranchi\& Matteucci 2003; Lanfranchi et al. 2006; ; Recchi et al. 2006; Marcolini et al. 2008).\\
Very recently,  Sawala et al. (2009) studied the chemical evolution of local dwarfs in a cosmological framework 
by means of high-resolution hydrodynamical numerical simulation, including also type Ia and type II SN enrichment. 
Their results indicate that the chemical properties of dwarf galaxies are driven by several aspects, such as the SN feedback, the
UV background and their gravitational potential. Their results stress the importance 
of efficient metal-enhanced outflows, fundamental for reproducing the stellar mass-metallicity relation. \\
A numerical study of the stellar metallicity distributions of dwarf spheroidals has been performed by Ripamonti et al. (2007). 
Their results point towards the substantial presence of very low metallicity [Fe/H]$<$-3 stars in dwarf spheroidals, in agreement 
with the results found in this paper. If no observational bias against these stars is present, possible 
solutions to this problem invoke a primordial IMF truncated for stellar masses below 1 $M_{\odot}$ or a pre-enrichment of the gas.  
However, very recently, the presence of extremely low metallicity stars has been detected  in ultrafaint dwarf galaxies (Frebel et al. 2009).

\begin{figure*}
\centering
\vspace{0.001cm}
\epsfig{file=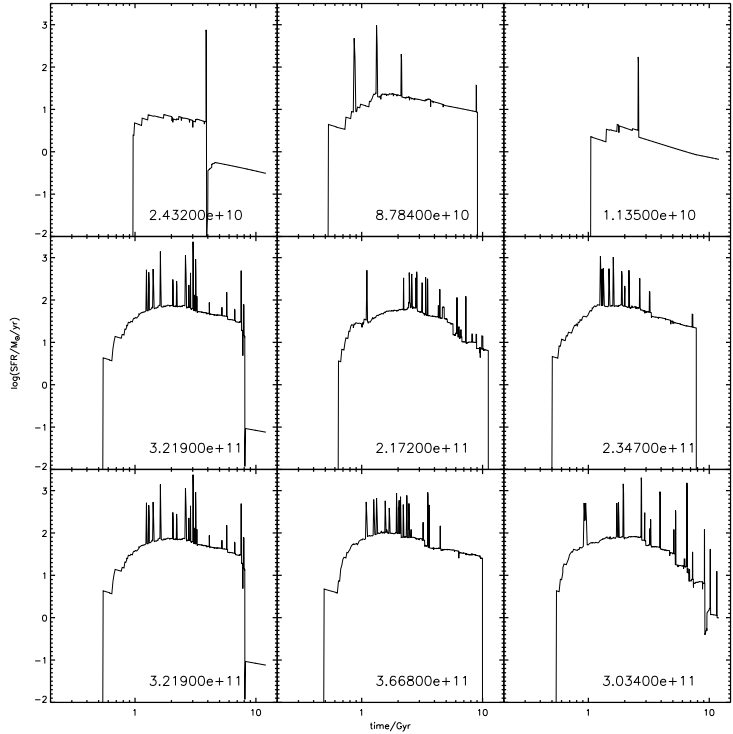,height=10cm,width=10cm}
\caption{ Star Formation histories of a few elliptical selected according the criteria described in 
Sect.~\ref{ell}.}
\label{sfr_ell}
\end{figure*}

\subsection{Abundances in elliptical galaxies}
\label{ell}
We conclude this section presenting our chemical evolution 
results obtained for elliptical galaxies. 
Elliptical galaxies are selected on the basis of their present-day (B-V) colour. 
In specific, we consider elliptical galaxies all the systems with present (B-V)$\ge 0.85$, 
following Roberts \& Haynes (1994).  
We perform a further cut on the basis of the present-day velocity dispersion $\sigma$,  
considering  all the systems with $30 \le log(\sigma/km/s)$. 
This range for the velocity dispersion is similar to the one spanned by the observational data of 
Thomas et al. (2005), who determined observationally 
the integrated stellar chemical abundances in a sample 
of local field ellipticals, and whose set of data will be used here for comparison 
with our predictions. 
In Fig.~\ref{sfr_ell}, we show the star formation histories of a few ellipticals selected 
according to the criteria described above. The numbers present in each panel of  Fig.~\ref{sfr_ell} indicate  
the present-day stellar masses of the selected galaxies. 
The selected SFRs are characterised by 
numerous peaks occurring in most of the cases 
at cosmic times  $\le 6$ Gyr, corresponding to redshift $z>1$. 
In general, galaxies with larger stellar masses 
exhibit higher SFR values.   
The SFHs of the objects shown in Fig. ~\ref{sfr_ell}, with 
little or no substantial SF in the last few Gyrs, 
are qualitatively in agreement with the SFHs of early type galaxies drawn from 
other SAMs (De Lucia et al. 2006; Pipino et al. 2009).\\
Generally, in local ellipticals the stellar abundances are
measured by means of absorption-line indices, such as
Mg$\, b$ and $<Fe> = 0.5 (Fe52720 + Fe5335)$ (see Thomas et
al. 2005 and references therein). This means that the observational abundances 
are luminosity-averaged values, which in principle represent underestimates to the 
true, mass-averaged abundances, which are the ones computed by means of our models, 
because of the fact that  at constant age, metal-poor stars are
brighter (Greggio 1997).  However, as discussed in several papers 
(Matteucci, Ponzone \& Gibson 1998; Thomas et al. 1999; Recchi et al. 2009), 
detailed chemical evolution calculations show that 
the difference between mass-averaged and luminosity-averaged abundances is very small, 
typically of the order of a few 0.01 dex. This result is not dependent on fundamental chemical 
evolution parameters, such as the stellar IMF (Recchi et al. 2009). For this reason, 
we neglect this difference and we compare our predicted 
mass-averaged stellar abundances to the ones derived observationally by means of 
absorption-line indices. 
%
\begin{figure*}
\centering
\vspace{0.001cm}
\epsfig{file=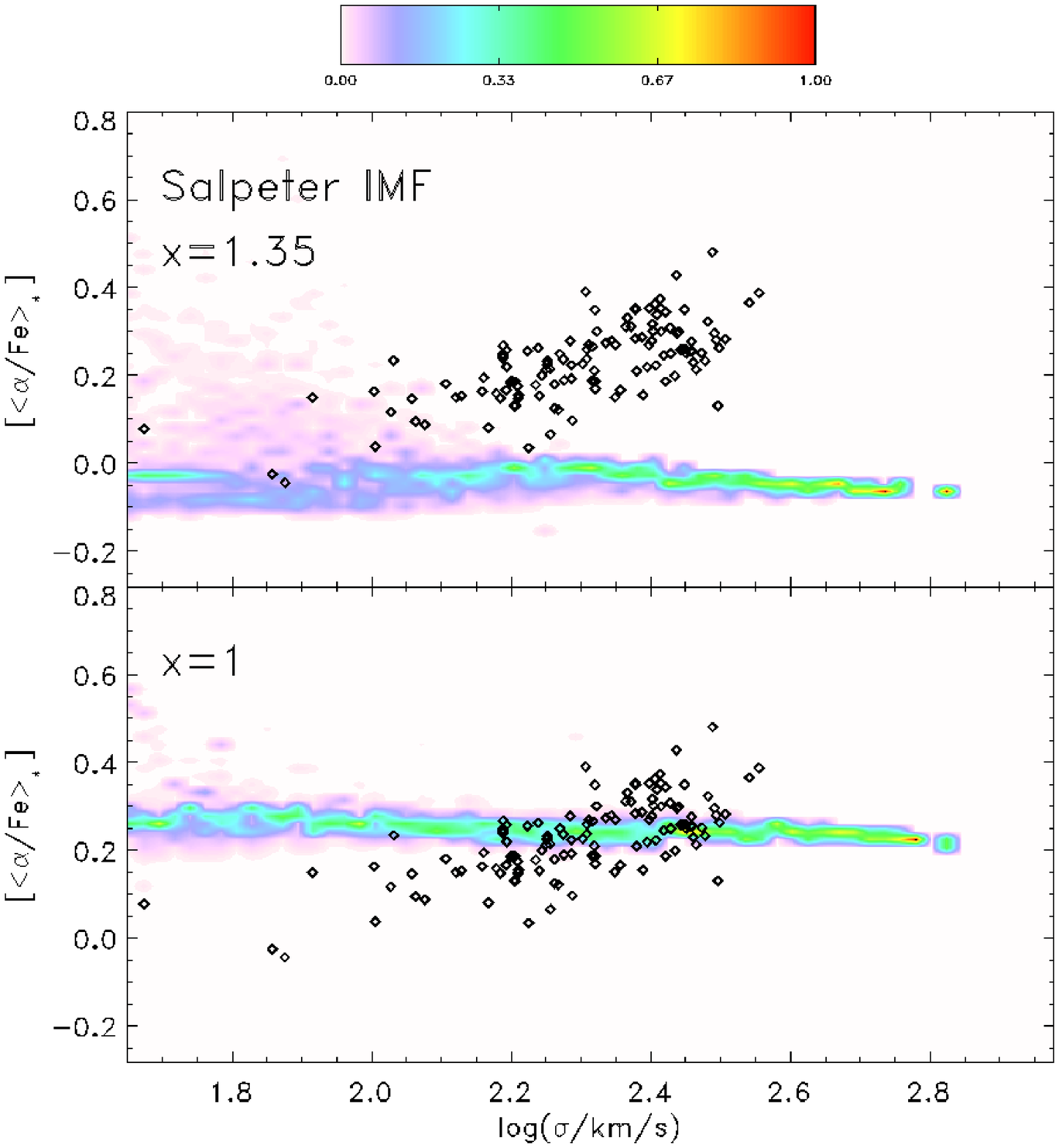,height=20cm,width=15cm}
\caption{ Predicted stellar average $\alpha/Fe$ ratio vs velocity dispersion  compared to local observations by Thomas et al. (2005) 
(open diamonds). 
The colour code, shown by the bar at the top of the Figure,  represents the predicted number of galaxies with given 
$[<\alpha/Fe>_{*}]$ with velocity dispersion $\sigma$, 
normalised to the total number of galaxies with that $\sigma$.
In the upper panel, we show our results computed with a Salpeter IMF 
whereas in the lower panel we have assumed an IMF slope of $x=1$. 
}
\label{alpha_fe_sigma}
\end{figure*}
\begin{figure*}
\centering
\vspace{0.001cm}
\epsfig{file=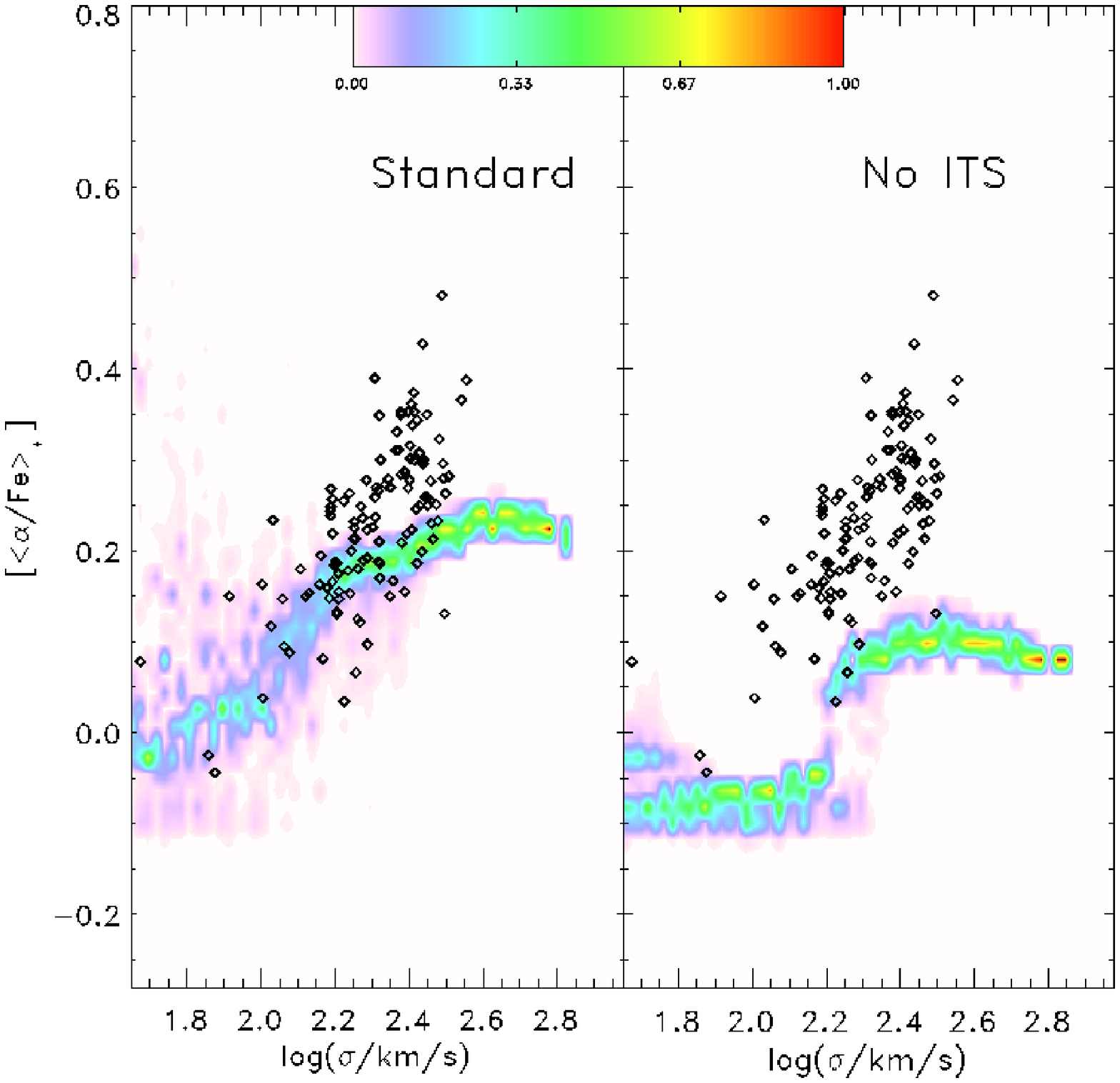,height=15cm,width=15cm}
\caption{ Predicted stellar average $\alpha/Fe$ ratio vs velocity dispersion  (contour), compared to local observations by Thomas et al. (2005) 
(open diamonds). The colour code represents the predictions, as described in Fig.~\ref{alpha_fe_sigma}. 
In the left panel, our results have been computed by means of a SF-dependent IMF and our standard assumptions, which include 
starburst-triggered interactions at high redshifts and AGN feedback. 
In the right panel, we have assumed a a SF-dependent IMF and AGN feedback but we have supressed starburst-triggered interactions.
}
\label{alpha_fe_sigma_noint}
\end{figure*}
\begin{figure*}
\centering
\vspace{0.001cm}
\epsfig{file=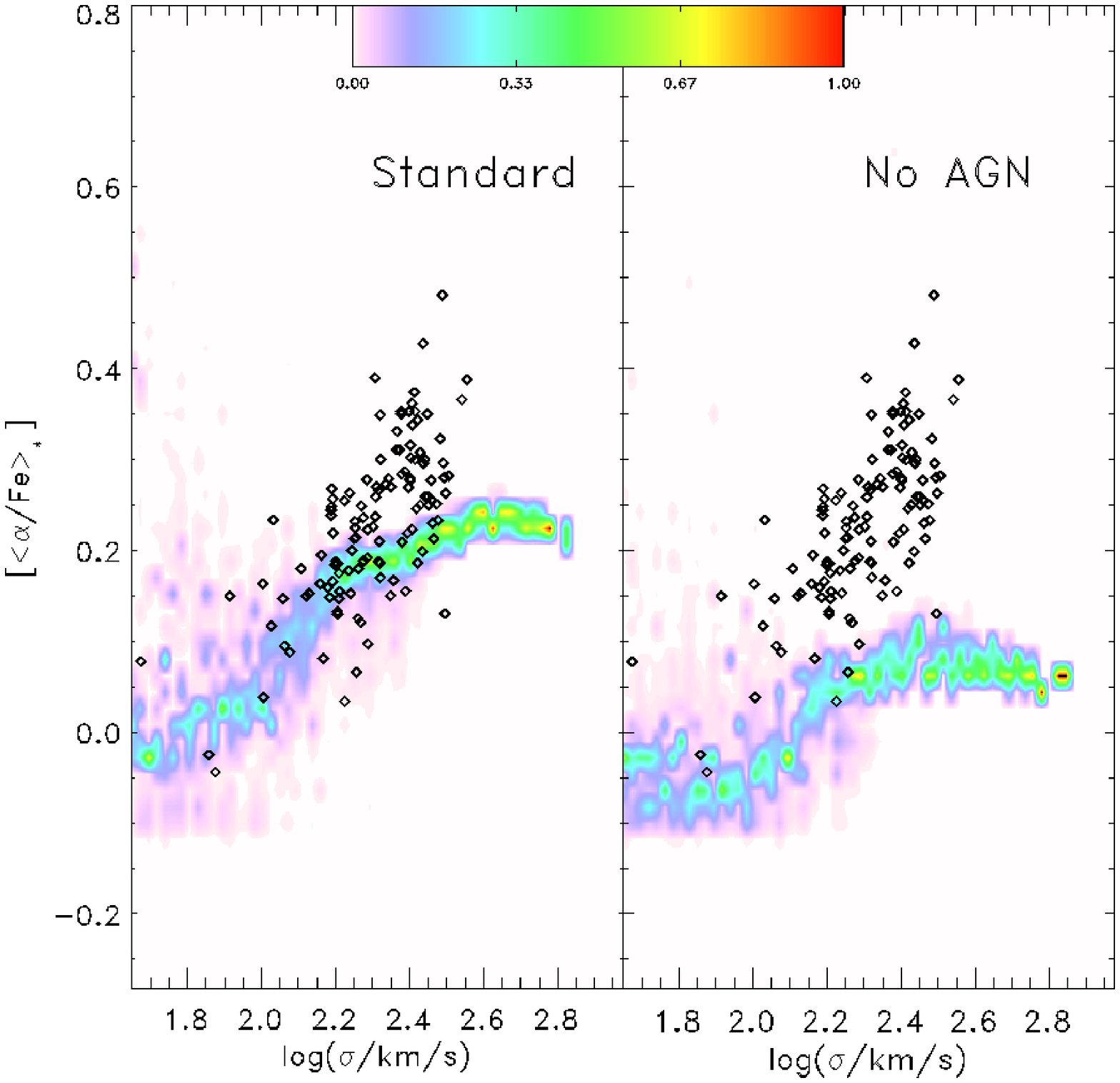,height=15cm,width=15cm}
\caption{ Predicted stellar average $\alpha/Fe$ ratio vs velocity dispersion  (contour), compared to local observations by Thomas et al. (2005) 
(open diamonds). The colour code represents the predictions, as described in Fig.~\ref{alpha_fe_sigma}. 
In the left panel, our results have been computed by means of a SF-dependent IMF and our standard assumptions, which include 
starburst-triggered interactions at high redshifts and AGN feedback. 
In the right panel, we have assumed a a SF-dependent IMF and starburst-triggered interactions 
but we have supressed  the effect of AGN feedback. }
\label{alpha_fe_sigma_noagn}
\end{figure*}

\begin{figure*}
\centering
\vspace{0.001cm}
\epsfig{file=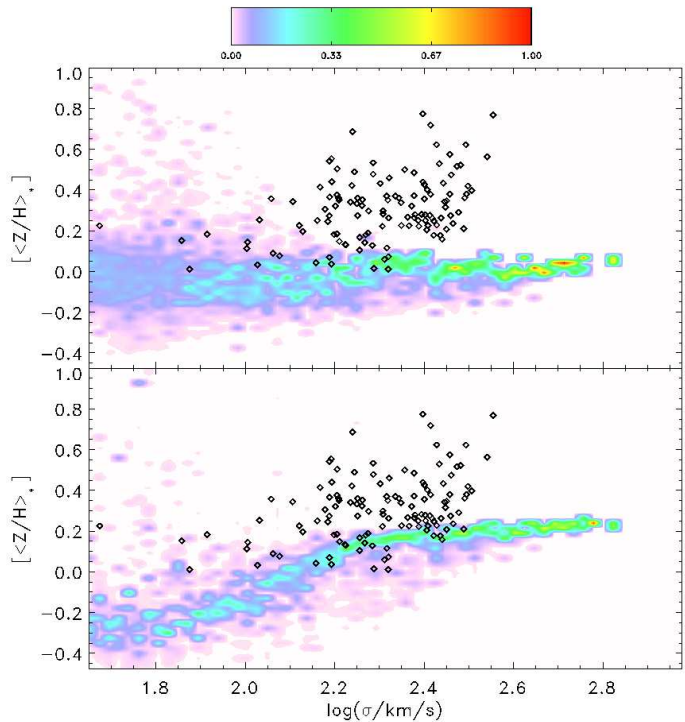,height=10cm,width=10cm}
\caption{ Predicted stellar average $Z/H$ vs velocity dispersion $\sigma$, compared to local observations by Thomas et al. (2005) 
(open diamonds). In the upper and lower panel, we show the results obtained by assuming a Salpeter IMF and a SF-dependent IMF 
(i.e. with $x=1.35$ for systems with SFR$\le 100 M_{\odot}/yr$ and with $x=1$ for systems with SFR$\le 100 M_{\odot}/yr$), respectively. 
The colour code, shown by the bar at the top of the Figure,  represents the predicted number of galaxies with given 
$[<Z/H>_{*}]$ with velocity dispersion $\sigma$, 
normalised to the total number of galaxies at the considered $\sigma$. }
\label{Z_sigma}
\end{figure*}

In the upper panel of Fig.~\ref{alpha_fe_sigma}, we show the predicted average stellar $\alpha$/Fe 
ratio versus velocity dispersion, computed assuming a Salpeter IMF and 
compared with a set of local observations by Thomas et al. (2005). 
In this case, the predicted $[<\alpha/Fe>_{*}]$ is computed considering O as representative 
of  $\alpha-$ elements. 
To compute our predictions, in Fig. ~\ref{alpha_fe_sigma} we use a standard Salpeter IMF, 
and a type Ia SN realization probability $A_{Ia}=0.002$. 
As we will see later, with our standard assumptions, this value 
allows us to correctly reproduce the zero point of the 
stellar [$\alpha$/Fe] - $\sigma$ relation. \\
While the  observations indicate a positive correlation between [$\alpha$/Fe] and $\sigma$, 
with galaxies with larger $\sigma$ presenting larger $\alpha-$enhancement, 
our predictions indicate a flat [$\alpha$/Fe] -$\sigma$ relation. 
In general, the observed  [$\alpha$/Fe] -$\sigma$ is interpreted in terms of smaller 
formation timescales in larger ellipticals, in which the chemical enrichment is dominated by the contribution 
of type II SNe,  hence most of the stars form with a chemical composition rich of  $\alpha$ elements (Matteucci 1994; Thomas et al. 2005). 
On the other hand, smaller galaxies present lower stellar [$\alpha$/Fe] since they form on longer timescales, comparable to the exploding 
times of type Ia SNe, whose contribution to the Fe enrichment of the ISM is relevant, lowering the average stellar [$\alpha$/Fe]. 
Matteucci (1994) has shown that a monolithical galaxy formation scenario for ellipticals is successful in reproducing 
the observed stellar abundance ratios. According to this scenario, 
in ellipticals star formation stops owing to the onset of galactic winds, which 
develop on smaller timescales in larger ellipticals, according to an inverse-wind scheme. 
Calura et al. (2008) have confirmed that gas outflows should occur on smaller timescales in large galaxies, according to 
a ``downsizing'' pattern for galaxy formation. \\
The impossibility to reproduce the observed [$\alpha$/Fe] -$\sigma$ relation is 
a well-known problem,  common to many hierarchical SAM. 
The reason is that, in hierarchical models, the star formation 
of large elliptical galaxies  is extended until relatively recent times, hence without avoiding that large amounts of Fe 
are restored by type Ia SNe into the ISM while star formation is still active. In general, this leads large galaxies to 
present lower average stellar 
[$\alpha$/Fe], i.e. to an anti-correlation between [$\alpha$/Fe] and $\sigma$  
(Thomas 1999; Nagashima et al. 2005; Pipino et al. 2009), contrary to what the observations indicate. 
On the other hand, by assuming a simple Salpeter IMF constant in time, our results indicate 
rather a flat relation 
between these two quantities. This is 
still at variance with the observations, however, this should be certainly regarded  as a
an improvement with respect to the previous attempts discussed above.\\
Our basic assuptions, i.e. a Salpeter IMF, the inclusion  of 
interaction-triggered starbursts and AGN feedback do not allow us to  
produce a positive correlation between the stellar [$\alpha$/Fe] and $\sigma$. 
In a recent paper, Arrigoni et al. (2009) studied the chemical evolution of local ellipticals 
by means of a semi-analytic model for galaxy formation which includes AGN feedback. 
Their main result concerns the capability to reproduce the observed correlation between  
stellar [$\alpha$/Fe] and  stellar mass, after a fine-tuning of the a type Ia SN constant
and by assuming an IMF slope $x=1.1$, where the canonical Salpeter slope is $x=1.35$. 
As already stressed, in chemical evolution studies the type Ia SN constant $A_{Ia}$ is an uncertain 
parameter, however the IMF is in general better constrained by local observations. 
In the light of the results, now we aim at studying the [$\alpha$/Fe] - $\sigma$ relation  
in the same conditions as the ones 
of Arrigoni et al. (2009), i.e. assuming a flatter IMF. 
In the lower panel of Fig. ~\ref{alpha_fe_sigma}, we show the stellar  
[$\alpha$/Fe]-$\sigma$ relation predicted for massive galaxies,  
obtained by considering a constant intial mass function with index $x=1$.  
This figure shows clearly that the assumption of a constant 
IMF flatter than the Salpeter has the effect of lifting the zero-point of the predicted 
[$\alpha$/Fe] -$\sigma$ relation, but has no effect on its slope. 
This is also in agreement with the discussion about the role of the IMF in the chemical evolution of early type galaxies 
of Pipino et al. (2009).  \\
Recent observational and theorethical results seem to indicate a dependency between the  
star-formation rate and the slope and the upper mass limit of the IMF.  
In normal star formation conditions, i.e. in local regions with SFRs $< 100 M_{\odot}/yr$,
a standard IMF slope $x \sim 1.3$ for  stellar masses $> 1 M_{\odot}$ 
is commonly accepted (Kroupa 2002, Recchi et al. 2009). 
On the other hand, in local starbursts, in the cores of stellar clusters and in ultra-compact galaxies, with SFRs $> 100 M_{\odot}/yr$, 
various results indicate a flatter IMF, with a slope $x \le 1$ (Elmegreen 2009; Dabringhausen et al. 2009).  
Further evidences in favour of a flat IMF in  environments undergoing strong star formation 
come 
from chemical evolution models, which, in order to be able to explain the oversolar [$\alpha$/Fe] values 
observed in the hot intracluster gas, must necessairily invoke an IMF skewed towards massive stars 
(Gibson \& Matteucci 1997, Portinari et al. 2004). 
Motivated by these results, now we aim at investigating  the impact of a SF-dependent IMF on the stellar 
[$\alpha$/Fe] vs $\sigma$ relation predicted by local ellipticals. 
Our aim is also to test how our results depend on other 
ingredients of the SAM model used in this paper, such as 
interaction-triggered starburst in massive galaxies and 
AGN feedback, and the importance of including or neglecting 
these effects 
in conjuction with the main chemical evolution parameter, i.e. the IMF. \\
In  the left panel of Fig. ~\ref{alpha_fe_sigma_noint}, we show the predicted 
[$\alpha$/Fe]-$\sigma$ relation obtained by considering a SF-dependent IMF, characterized by 
a standard Salpeter slope ($x=1.35$) in objects with SFRs $< 100 M_{\odot}/yr$, and 
$x=1$ in objects with SFRs $> 100 M_{\odot}/yr$. This assumption in consistent with the study of Dabringhausen 
et al. (2009), indicating IMF slopes $x \le 0.7$ in starburst and ultra-compact galaxies, 
with SFRs of $10- 100 M_{\odot}/yr$. Here we consider 
an IMF slope $x =1$, 
close to the upper limit of Dabringhausen et al. (2009), since this does not require 
substantial modifications of the parameters of our SAM to reproduce  the main local observational constraints, such 
as the luminosity function or the Tully-Fisher relation.  
In this case, we see that the use of this IMF allows us to considerably improve our results, 
reproducing the observed correlation between 
[$\alpha$/Fe]-$\sigma$ and 
bringing our results in good agreement with the  
observations on a large range of velocity dispersions. 
In the right panel of Fig. ~\ref{alpha_fe_sigma_noint}, we show our results obtained with a model 
which does not include encounter-triggered
starbursts, but it does include a SF-varying IMF as described above. 
In this case, the  predicted [$\alpha$/Fe]-$\sigma$ relation is too shallow to 
reproduce the observations. 
This result outlines the importance of interaction-driven starbursts in determining the 
[$\alpha$/Fe]-$\sigma$ relation of local elipticals. 
At this stage, it may be interesting to test also the role of the AGN feedback in the chemical 
evolution of ellipticals. 
In Fig.~\ref{alpha_fe_sigma_noagn}, we compare our results obtained with our standard model and with a SF-dependent IMF with the ones 
achieved with 
a model which does include interaction-triggered starbursts, but which does not take into account the effect 
of AGN feedback in massive galaxies. 
The model with no AGN feedback has a shallower slope than the standard model and produces a poor match to the observations. 
This indicates that also the AGN feedback must play some important 
role in shaping the [$\alpha$/Fe]-$\sigma$ relation of local ellipticals. \\
In summary, in order 
to reproduce the correlation between the stellar [$\alpha$/Fe] and $\sigma$ of local ellipticals with our SAM, three basic 
ingredients seem to be required: (i) encounter-triggered starbursts at high redshift; (ii) AGN feedback; (iii) a SF-dependent IMF, 
flatter than the Salpeter in systems with strong SF. 
It is worth to note how a small decrease of the IMF slope allows us 
to considerably  improve our results. 
A further decrease of the IMF index $x$ may possibly improve our results, but, in order to reproduce 
the local constraints, 
it would probably require 
a re-calibration  of the main parameters of the SAM to reproduce the local constraints. This will be the subject of future work. \\
The predicted mass-stellar metallicity relation  is shown in 
Fig. ~\ref{Z_sigma}, compared to the same relation as observed in local ellipticals (Thomas et al. 2005). 
In the upper panel of Fig. ~\ref{Z_sigma}, we present our results 
obtained with a Salpeter IMF. In this case,  
we predict a basically flat 
$[<Z/H>_*]$ vs $\sigma$ relation, whereas the data indicate a positive  
correlation between these two quantities. 
In the lower  panel of Fig. ~\ref{Z_sigma}, the 
$[<Z/H>_*]$ -$\sigma$ relation is computed by assuming a SF-dependent IMF as described above. 
Again, this assumption improves considerably our fit to the observational data. This plot povides a further, independent 
indication that the assumption of a SF-dependent IMF is a possible solution to the long-standing problem 
of explaining the chemical abundances in ellipticals with hierarchical SAMs. \\ 
When comparing our predictions with observations, 
it is worth to stress that the observational $[<Z/H>_*]$ vs $\sigma$ 
is likely  affected by a 
template bias (Proctor et al. 2004; Thomas et al. 2005; Nagashima et al. 2005), 
which reflects the fact that the stellar population models used to determine the metallicities are based on stellar spectra 
in which some stars present a sub-solar [$\alpha$/Fe]. 
This effect could cause a steep $[<Z/H>_*]$ vs $\sigma$ relation, but, as shown by Thomas et al. (2005), it leaves basically unchanged the 
$[<\alpha/Fe>_*]$  vs $\sigma$ relation. 
Furthermore, as outlined by Pipino et al. (2009), the fact that 
elliptical galaxies  exhibit strong metallicity gradients makes a detailed 
comparison between our predicted $[<Z/H>_*]$ vs $\sigma$ relation with observations quite difficult, 
and is likely to mimic a steeper $[<Z/H>_*]$ vs $\sigma$ relation, leaving essentially unchanged the $[<\alpha/Fe>_*]$  vs $\sigma$. \\
Finally, it is important to stress that in this paper, to compute the integrated stellar abundances in 
early-type galaxies, 
we have considered the 
integral SF histories of our selected  galaxies, represented by the sum at each timestep of the SF  histories 
of all the progenitors. 
As stressed by Pipino et al. (2009), 
the most massive progenitors have in general shorter star formation timescales and should dominate the 
integrated stellar abundances. 
For some systems, this may  cause an underestimation of  the  
stellar  [$\alpha$/Fe] ratios by 0.1-0.2 dex. 
This effect is not taken into account at the present time, as considering in detail 
the star formation histories of the single progenitors for all our galaxies would require considerable computational times. 
A subsequent paper completely dedicated to this aspect is currently 
under preparation. 

\section{Discussion and conclusions}
~\label{concl}
The major aim of the present paper is to 
model chemical evolution by means of a hierarchical SAM by 
taking into account properly all the various stellar sources. 
We also aim at providing 
chemical evolution predictions 
computed by means of star formation and infall histories 
derived from first principles, by means of an ab-initio approach.  
Beside high-mass stars, dying as core-collapse type II SNe, 
we consider also the contributions from type Ia SNe and low and intermediate mass stars. 
A continuous delay-time distribution is assumed for type Ia SNe, 
relevant producers of some important elements, such as Fe and Si. 
We also consider finite lifetimes for low and intermediate mass stars, 
which are important producers of C and N. Our methods allow us to 
provide, for the first time, chemical evolution predictions for a large set of chemical 
elements, produced by stars of various types on different 
timescales. This is particularly useful to 
study the abundance ratios between elements synthesised by different stellar sources 
on various timescales, which are helpful tools to measure the timescales of star formation 
in galaxies of various masses and morphologies.  
We studied chemical evolution in various galactic types, comparing our predictions with available 
observations of chemical abundances in the Milky Way, in local dwarf galaxies and in local ellipticals.\\
First, we studied the abundance ratios vs metallicity for Milky Way-like galaxies. 
We 
compared our predictions to observations available for local stars of the Solar Neighbourhood 
for various elements. 
Overall, concerning the most important constraints, we found a good agreement 
between our predictions and the observations.
In the [O/Fe] vs [Fe/H] plot, our results for MW-like galaxies indicate 
the presence of 
a horizontal population of stars with metallicity $-2 \le$ [Fe/H] $\le -0.2$ and 
constant [O/Fe]$\sim -0.2$- $-0.3$.  This population is the result of 
substantial 
late infall episodes of pristine gas, 
very frequent in the selected MW-like galaxies, which 
have the effect of decreasing the 
metallicity of the most recently formed stellar populations, leaving unchanged 
their [O/Fe] ratio. 
When we consider only the galaxies with
at least 75 \% of the present-day mass already assembled at $z=2$, 
this horizontal population of stars with constant [O/Fe]$\sim -0.3$ disappears. 
These galaxies do not experience major 
accretion episodes at epochs after $\sim 6$ Gyr. 
If we consider only these subset of galaxies, 
the observational 
[O/Fe]-[Fe/H] is accounted for. 
Our results concerning the [O/Fe] vs [Fe/H] plots 
exclude that 
our Galaxy experienced major gas accretion episodes within the last few Gyrs, either due to merging with 
gas-rich, low-metallicity dwarf galaxies, or due to infall of pristine gas. 
This is 
is in agreement with a large set of previous results based on the
interpretation of abundances observed in MW disc and halo 
stellar populations and on kinemathical studies of the various components of the MW (e.g. Gilmore et al. 2002; Colavitti et al. 
2008 and references therein). \\
The typical decrease of the observed [$\alpha$/Fe] vs [Fe/H] plot 
is reproduced, as well as the observed trends for the other abundance ratios vs [Fe/H]. 
However, for other elements, the predictions are shifted 
downwards with respect to the observed abundance ratios. This is due to the adopted stellar yields, which 
in most cases lead to underestimating the observed abundance pattern. 
For elements other than O, 
the uncertainty in the stellar yield is a well-known problem of chemical evolution models of the solar neighbourhood (Fran\c cois et al. 2004). 
Also the IMF is likely to play some effects on the [X/Fe] vs [Fe/H] plots. 
For instance, with a Salpeter IMF, we predict an anti-corrlation between  [C/Fe] and [Fe/H], whereas the observational 
data indicate a substantially flat behaviour. 
We have shown how, by assuming  the IMF of Kroupa et al. (1993) and by slightly modifying the stellar yields of C 
for both intermediate mass- and massive stars, it is possible to improve our fit to the  [C/Fe] - [Fe] diagram. 
From the study of the local 
stellar metallicity distribution, we showed that we can correctly reproduce 
the position of the peak and the low-metallicity tail of the observed SMD. 
Also the predicted fraction of stars with metallicities $0\le [Fe/H]\le +0.2$ is in reasonable agreement with the observational data.\\
By assuming that the chemical composition of the outflows is the same as the one of the ISM and the same value for  
type Ia SN realization probability  $A_{Ia}$ as assumed for MW-like galaxies, 
we severely overestimate the stellar metallicities of dwarf galaxies. 
Differential outflows, where metals are lost more efficiently than H and He, have a major effect 
on the slope of the predicted $<[Fe/H]>_* - M_{*}$ relation, whereas the quantity 
$A_{Ia}$ affects primarily the zero point of this relation. 
The assumptions of strongly enhanced outflows and of a lower value for $A_{Ia}$ 
substantially decrease the 
the average stellar metallicity of dwarf galaxies, 
yielding a better match between predictions and observations in the $<[Fe/H]>_* - M_{*}$ diagram. 
Other theoretical results outline the role of metal-enhanced outflows in dwarf galaxies (see Recchi et al. 2008). 
The physical justification for differential mass loss from dwarf galaxies 
has been studied in the past by means of numerical simulations. 
MacLow \& Ferrara (1999) modelled 
the effects of SN explosions in dwarf galaxies, finding that the metal ejection is much more 
efficient in lower mass galaxies. 
By means of 2D hydrodynamical simulations, D'Ercole \& Brighenti (1999) studied the feedback of a starburst on the ISM of typical gas-rich dwarf galaxies.
They showed that metals are expelled more easily than the global ISM since  
the metal-rich material shed by the massive stars belongs to a very hot phase of the ISM, which can be easily accelerated to velocities higher than 
the escape speed and leave the galaxy. 
On the other hand, the interstellar gas, heated up by 
the SN explosions is only temporarily affected by the starburst, and the galaxy is able to recover a cold 
ISM after a time of the order of $\sim$ 100 Myr from the starburst. 
Also results from chemical evolution models suggest that heavy elements should be lost more effectively than H
to reproduce the global properties of dwarf galaxies (Pilyugin 1993, Marconi et al. 1994). 
The existence of metal-enhanced outflows  
from star-forming galaxies has also been recently confirmed by observations of local starbursts (Martin et al. 2002; Ott et al. 2005). \\
Our results concerning the abundance ratios of dwarf galaxies are compared to the 
abundances observed in local dSphs. Also in this case, 
the characteristic 
decreasing trend of [$\alpha$/Fe] vs [Fe/H] is reproduced, and explained 
as due to the delay between type II SNe explosions and type Ia 
SN explosions. \\
The study of the individual and cumulative 
stellar metallicity distributions of dwarf galaxies 
of various masses  provides a satisfactory agreement between predictions and observations. 
Also this result is a direct consequence of having assumed metal enhanced outflows in dwarf galaxies and 
a low value for $A_{Ia}$. The only discrepancy concerns the fraction of stars with metallicity [Fe/H]$\le -2$. Our predictions 
indicate that they must be present in substantial fractions, in particular in galaxies with the lowest masses. 
The observational SMDs used in this paper 
do not indicate the presence of these low metallicity stars in dwarf spheroidals. 
However, 
very recently a few extremely low metallicity stars 
have been detected  in the Sextans and Draco dSphs and in ultrafaint dwarf galaxies 
(Aoki et al. 2009; Cohen \& Huang 2009;  Frebel et al. 2009).\\
Finally, we studied the  chemical evolution of local ellipticals. 
Our results were compared with  
the available observations, which represent the average stellar 
abundances in local early-type galaxies. 
Concerning the [$\alpha$/Fe] vs $\sigma$ plot, 
the observations indicate a positive correlation between [$\alpha$/Fe] and $\sigma$, usually interpreted 
in terms of more efficient 
star formation efficiency in larger galaxies (Matteucci 1994; Thomas et al. 2005). 
Previous attempts to reproduce the observations  in early type galaxies by 
means of semi-analytical $\Lambda$CDM galaxy formation models have led to several insuccesses. 
Theoretical results by various authors point  towards an anti-correlation between $\alpha$/Fe and $\sigma$
(Thomas 1999, Nagashima et al. 2005).  \\
We have performed a detailed study of the role of various ingredients in determining the slope and 
the zero point of the $\alpha$/Fe  - $\sigma$ relation. 
Our predictions computed with a constant Salpeter IMF    
indicate a flat [$\alpha$/Fe] -$\sigma$. 
We have found that, by assuming a SF-dependent IMF, 
Salpeter-like (i.e. with slope $x=1.35$) in 
objects with SFR $< 100 M_{\odot}/yr$ and slightly flatter (with $x=1$) in systems with strong star formation, 
the observed correlation can be accounted for, at least on the velocity dispersion range $1.8 \le log (\sigma/kms/s) \le 2.2$. 
A major role is played by starburst-triggered interactions and AGN feedback. 
In fact, we have seen that when we relax any of these two ingredients in our models, we predict a 
flatter [$\alpha$/Fe] -$\sigma$ relation than the one produced considering both effects and the SF-dependent IMF. 
Our main result is then that, in order to reproduce this correlation, all these three ingredients must be taken into account. 
The discrepancy between our 
predicted [$\alpha$/Fe] -$\sigma$ relation and the observations for $log (\sigma/kms/s) > 2.3$ may be due to the fact that, at the present stage, 
the integrated stellar [$\alpha$/Fe] have been computed by considering the 
integral SF histories of early type galaxies, and not the contributions of the single progenitors. 
Since in general the most massive progenitors should exhibit the highest stellar [$\alpha$/Fe] and have a higher weight in the 
calculation of the total [$\alpha$/Fe], a more detailed computation of the stellar 
[$\alpha$/Fe] values taking into account each single progenitor could further steepen the predicted [$\alpha$/Fe]-$\sigma$ relation 
and alleviate the discreancy for the largest mass systems. 
This task requires 
long computational times and will be the subject of a forthcoming paper. 
With a standard Salpeter IMF, we cannot reproduce 
the slope of the 
mass-stellar metallicity relation observed  in ellipticals, 
which should however be 
considered a less robust observational constraint, mainly owing to a possible 
template bias affecting the stellar population models used to determine the metallicities. 
However, we have shown that, also in this case, 
 the assumption of a SF-dependent IMF alleviates the discrepancy between our predictions and the observational data.\\
In the future, our 
semi-analytical model for galaxy formation will be used to perform a more thorough 
study of the abundance ratios in ellipticals and of the mass-metallicity relation 
and its evolution (Calura et al. 2009). 

\section*{Acknowledgments}
We wish to thank S. Recchi, G. Cescutti, F. Matteucci and G. De Lucia for 
several stimulating discussions. 
G. Lanfranchi and E. Colavitti are acknowledged for providing us with compilations of observational 
data for the Milky Way and for local dwarf spheroidal galaxies.
An anonymous referee is acknowledged for useful suggestions and for a careful reading of the paper. 
This work was partially supported by the Italian Space Agency
through contract ASI-INAF I/016/07/0.
F.C. acknowledges financial support from PRIN2007, Prot.2007JJC53X\_001.

\label{lastpage}

\end{document}